\documentclass[12pt]{article}
\pdfoutput=1

\usepackage{jheppub}
\usepackage{amssymb,amsfonts,amsmath}
\usepackage{tikz-cd}
\usepackage{mathtools}
\usepackage{amsfonts}
\usepackage{enumerate}
\usepackage{mathrsfs}
\newcommand{\boundellipse}[3]
{(#1) ellipse (#2 and #3)
}
\usepackage{ textcomp }
\usepackage{tikz}
\usepackage[section]{placeins}
\usepackage{verbatim}
\usepackage{amsthm}
\usepackage{tikz}
\usetikzlibrary{shapes,arrows,chains}
\usetikzlibrary{decorations.markings}
\usetikzlibrary{decorations.pathmorphing}
\usetikzlibrary{shapes.multipart}
\tikzset{snake it/.style={decorate, decoration=snake}}
\usepackage{url}

\allowdisplaybreaks

\newcommand\rH{\mathrm H}
\newcommand\bC{\mathbb{C}}
\newcommand\rU{\mathrm{U}}
\newcommand\cA{\mathcal A}

\newcommand\bZ{\mathbb Z}
\newcommand\cC{\mathcal C}
\newcommand\rB{\mathrm B}

\newcommand\bF{\mathbb F}

\newcommand{\rC}{\mathrm{C}}
\newcommand\pt{\mathrm{pt}}
\renewcommand\d{\mathrm{d}}
\DeclareMathOperator{\Sq}{Sq}
\renewcommand\H{\operatorname{H}}

\DeclareMathOperator{\SH}{SH}

\DeclareMathOperator{\Spin}{Spin}
\DeclareMathOperator{\inv}{inv}
\DeclareMathOperator\Ext{Ext}

\DeclareMathOperator{\fGP}{fGP}
\DeclareMathOperator{\Sym}{Sym}

\newcommand\longto\longrightarrow
\newcommand\longfrom\longleftarrow

\DeclareMathOperator{\im}{im}

\DeclareMathOperator{\ob}{ob}
\DeclareMathOperator{\Hom}{Hom}
\usepackage{xcolor}

\newcommand{\be}{\begin{equation}}
\newcommand{\ee}{\end{equation}}
\newcommand{\bea}{\begin{eqnarray}}
\newcommand{\eea}{\end{eqnarray}}
\DeclareMathOperator{\spec}{Spec}
\DeclareMathOperator{\Tor}{Tor}

\newcommand{\bra}[1]{\langle #1|}
\newcommand{\ket}[1]{|#1\rangle}

\newcommand{\cZ}{\mathcal{Z}}
\newcommand{\CA}{\mathcal{A}}

\newcommand{\CZ}{\mathcal{Z}}

\newcommand\qt\tau

\newcommand{\p}{\partial}

\renewcommand{\tilde}[1]{\widetilde{#1}}

\makeatletter
\renewcommand{\@seccntformat}[1]{\csname the#1\endcsname.\,\,}
\makeatother
\allowdisplaybreaks

\let \savenumberline \numberline
\def \numberline#1{\savenumberline{#1.}}

\makeatletter
\def\@fpheader{\relax}
\makeatother

\def\bea{\begin{eqnarray}}
\def\eea{\end{eqnarray}}

\title{\ \vspace{1.6cm} \\
\scalebox{0.9}{Symmetries and Anomalies of (1+1)$d$ Theories: }
\scalebox{0.9}{
2-groups and Symmetry Fractionalization}}
\author{
Matthew Yu
}

\emailAdd{myu@perimeterinstitute.ca}
\affiliation{
        Perimeter Institute for Theoretical Physics\\
        31 Caroline St N, Waterloo, ON N2L 2Y5, Canada}
\abstract{We investigate the interactions of discrete zero-form and one-form global symmetries in (1+1)$d$ theories.  Focus is put on the interactions that the symmetries can have on each other, which in this low dimension result in 2-group symmetries or symmetry fractionalization.  A large part of the discussion will be to understand a major feature in (1+1)$d$: the multiple sectors into which a theory decomposes.  We perform gauging of the one-form symmetry, and remark on the effects this has on our theories, especially in the case when there is a global 2-group symmetry. We also implement the spectral sequence to calculate anomalies for the 2-group theories and symmetry fractionalized theory in the bosonic and fermionic cases.  Lastly, we discuss topological manipulations on the operators which implement the symmetries, and draw insights on the (1+1)$d$ effects of such manipulations by coupling to a bulk (2+1)$d$ theory. }

\begin{document}

\maketitle
\newpage 
\section{Introduction}
Understanding symmetries is key to revealing many nontrivial features of quantum field theories.  In different dimensions, various higher form symmetries may exist in a theory.  These symmetries can be encoded in the higher codimension operators, and in many cases have the structure of a higher category.  Global symmetries do not need to stay stagnant either: one can promote the symmetry to a dynamical symmetry under the assumption of the symmetry having no t'Hooft anomaly.  This is done by coupling to a background connection and gauging the symmetry.  For finite symmetries, another modern point of view of gauging is performing a categorical condensation \cite{Gaiotto:2019xmp}.  Furthermore, the symmetries can interact with each other in nontrivial ways to give higher $n$-groups.  It has therefore become increasingly necessary to implement techniques from category theory and topology to consolidate information about symmetries.  This includes, the possible 't Hooft anomalies, the algebraic structure of higher codimensional defects, and the groupoid of ways in which theories are related to each other by gauging \cite{Gaiotto:2020iye,Gaiotto:2017yup,Hsin:2020nts}.

The purpose of this paper is to study theories in (1+1)$d$ that exhibit a discrete global zero-form and one-form symmetry.  We explore two possibilities: the first is when the zero-form symmetry is nontrivially extended by the one-form symmetry leading to a 2-group, and the other is when the extension is trivialized leading to symmetry fractionalization. Theories in (1+1)$d$ are interesting from this point of view because the higher form symmetries are restricted only to one-forms, and it is possible to keep track of the interactions between the line operators that implement the one-form symmetry and the point operators that implement the zero-form symmetry.  It is also possible to calculate the anomalies of theories with 2-group symmetries and theories with symmetry fractionalization rigorously in this low dimension. We present explicit calculations and give formulas for the cohomology that classify the anomalies.

An aspect that we will emphasize is the notion of having multiple ground states, or disjoint sectors, in a theory. In (1+1)$d$ theories, the one-form symmetry gives information about the number of local ground states. Not only is there information about the multiple subsectors in a theory, but there is also information contained in moving between the sectors.  We address these two points by laying down some theoretical framework and also giving examples. 
Furthermore, we discuss gauging one-form symmetries and explaining the dual $(-1)$-form symmetry from a physical and mathematical viewpoint.  We find that if we gauge the one-form group in a 2-group, the extension in the 2-group becomes a mixed anomaly that restricts to each subsector.  It is also natural to apply this knowledge of subsectors on the side of symmetry fractionalization,  from which it is possible to make relations with discrete torsion.

The layout of the paper is as follows: in \S\ref{startoftwo} we begin by reviewing generalized symmetries and give a precise mathematical definition of one-form symmetry and of 2-groups. We then focus on gauging in a theory with 2-group symmetry, outlining properties of one-form symmetries as also discussed in \cite{Sharpe:2019ddn}.  We also track the relationship between mixed anomalies and extensions, given by the Serre spectral sequence, at the level of partition functions. In \S\ref{Bosonic2GAnomaly} and \S\ref{fermioic2GAnomaly} we employ spectral sequence techniques to calculate anomalies for (1+1)$d$ bosonic and fermionic theories, respectively.  We discuss symmetry fractionalization in \S\ref{symfracsection} and relate it to discrete torsion for the zero-form symmetry, a special feature that exists in (1+1)-dimensions.  We then give the anomalies for (1+1)$d$ theories exhibiting symmetry fractionalization.  Finally, we finish off by discussing manipulations regarding the topological sectors of (1+1)$d$ theories and how they can be recovered by coupling to a bulk topological field theory (TFT) \cite{Kapustin:2014gua}.

\section{2-group Global Symmetry in (1+1)d Theories}\label{startoftwo}
Given a $d$-dimensional quantum field theory with a $p$-form global symmetry, one is also provided with a particular set of topological codimension $p+1$ operators: the charged operators for the $p$-form symmetry, which implement the group action of the $p$-form symmetry upon crossing these operators \cite{generalizedSymmetries}.  By virtue of their group-structure, these symmetry defects form a subset of all invertible topological operators in the theory. Another set which exists is the collection of non-invertible defects \cite{Chang:2018iay}, which we will not consider for our purposes.  The fact that the group like property of higher form symmetries are encoded in a collection of invertible defects of various codimensions implies that in (1+1)-dimensions the only higher form symmetry that is possible is a one-form symmetry.  While zero-form symmetries of a theory are well understood to be described by ordinary groups, a generalization is required to talk precisely about the group structure of one-form symmetries and 2-groups;  this will be the focus of the following subsections.  

\subsection{Defining a 2-group: 1-forms}\label{defining2groups}
We will build to the definition of a 2-group by introducing some formal definitions from category theory and homotopy theory required to sharply define a 2-group.   We will also illustrate how 2-group symmetries are a group of one-form symmetries and zero-form symmetries ``interacting'' with each other.  Along the way we will have properly defined the notion of one-form symmetry, and the analogue applies for higher form symmetries in higher dimensions. 
%

We start off with the notion of a \textit{group object}, which is a generalization of the structure of groups to categories other than $\mathbf{Set}$, the category of sets. That is to say, the underlying set of elements for the group, which is typically an object of $\mathbf{Set}$, is replaced by another object from some other category.  More formally, if $X$ is a group object in the category $\mathcal{C}$, then there are maps:
\begin{equation}
    m: X \times X \to X\,, \quad e: 1 \to X\,, \quad \inv: X \to X,
\end{equation}
where $m$ is an associative multiplication, $e$ is a map from the terminal object $1 \in \mathcal{C }$ which is a two sided unit of $m$, and $\inv$ is the two sided inverse of $m$ \cite{lane2013categories}.  We see from the above requirement that one recovers the traditional notion of a group if $\mathcal{C} = \mathbf{Set}$ and $X$ is a group object in $\mathbf{Set}$. In this case we define $m$ such that it takes the form of group multiplication for the underlying set, $e$ such that it selects the identity of $X$, and $\inv$ such that it assigns to all group elements its inverse.  Hence, one could say the collection of zero-form symmetries of a theory is described by a group object in the category $\mathbf{Set}$.

We now turn to a generalization of groups: \textit{groupoids}.  A groupoid has the features of a group in which any two elements may not be meaningfully composed.  In particular, a groupoid is a (small) category where every morphism is an isomorphism. The category of all groupoids is $\mathbf{Grpd}$.  The objects in this category are groupoids, which are categories themselves, and the morphisms between objects are actually functors of groupoids.  We can simplify this generalization to once again recover the traditional notion of a group, which is given by the \textit{morphisms in a groupoid}.

Suppose that a groupoid only has one object, and consider a group $G$.  We could have a morphism from that one object to itself, with each morphism given by an element $g\in G$. Any two morphisms given by $g_1,g_2 \in G$ may be composed, just as in the group.  Associativity of morphisms holds because of the associativity of the group operation, and the identity of $G$ is the identity morphism for the object of the groupoid. Hence, the morphisms of a groupoid with only one object form a group under composition.  This is a notion of \textit{delooping} applied to a group, which categorifies it into a groupoid.

Returning to physics, we say a ``group of 1-form symmetries'', denoted $\cA_{[1]}$, is a group object in the category of groupoids with only one object.  To reflect this, we introduce the following notation:
\begin{align}\label{1formgroup}
    \pi_0 \, \mathcal{A}_{[1]} &:= \ob(\CA_{[1]}) = \{\ast\}\,\notag \\
    \pi_1 \, \mathcal{A}_{[1]} &:= \Hom(\ast, \ast) = A\,,
\end{align}
where $\{\ast\}$ denotes the single object of the groupoid. Here the group of morphisms $\pi_1\,\CA_{[1]}$ is what is meant when a said theory possesses an ``$A$ group one-form symmetry".  The composition of maps also endows $\mathcal{A}_{[1]}$ with a unique group law. Notably, as a groupoid with group law, $\mathcal{A}_{[1]}$ contains a multiplication $m:\mathcal{A}_{[1]} \times \mathcal{A}_{[1]} \rightarrow \mathcal{A}_{[1]} $, and an associator $\beta$.  The associator is a 3 cocycle such that for $a_1, a_2, a_3 \in \ob(\mathcal{A}_{[1]})$, $\beta(a_1,a_2,a_3)$ gives the isomorphism $m(m(a_1,a_2),a_3) \simeq m(a_1,m(a_2,a_3))$ for the associativity of $m$. An important point to make is that in order for this group law to be unique, we require $A$ to be abelian.   Henceforth, when we mention the specific one-form symmetry group, what one should think of is the underlying $A$. This is what is really meant when thinking of one-form symmetries as groups in the regular sense, such as the center $\mathbb{Z}_N$ one-form symmetry in 4$d$ $SU(N)$ pure Yang-Mills theory. 

For the purpose of this paper we will take $A$ to be a finite group, and in later sections we will consider more specific cases for this finite group.
In the subsequent sections, we could also refer to the one-form symmetry as $\rB A$.  Analogously,  $\rB G$ where $G$ is treated as a $(p-1)$-form group can be thought of as $K( G, p)$, the $p$-th Eilenberg-Maclane space  of $G$.  When computing group (super)cohomology in the later sections we are computing (super)cohomology on the classifying space of the group, rather than cohomology of the topological space of the group.  In our case, the latter would be a finite set of points, which does not have interesting cohomology.  In order to conserve on notation we will write $\rH^\bullet(G)=\rH^\bullet_{group}(G) = \rH^\bullet(\rB G)$, to reflect this fact for $G$ the zero-form symmetry.   For the one-form symmetry, we will denote the cohomology by $\rH^\bullet(\mathcal{A}_{[1]})$ or $\rH^\bullet(\rB A)$, to mean the cohomology of the space $\rB A$.  We follow the usual slight abuse of notation, taking the space $\rB A$ as the space that carries a unique group structure that is the underlying group for a 1-form symmetry, and also taking $\rB G$ to be the delooped space which gives the group cohomology of $G$.

\subsection{Defining a 2-group: Including 0-form}

In general, it is possible to have zero-form symmetries along with one-form symmetries; we will take the zero-form symmetries to be the group $G_{[0]} = G$ which is a finite group.  Furthermore, there could be a nontrivial action of the zero-form on the one-form symmetry.  Our goal will be to study groupoids with group law, $\mathbb{G}$, that fit in a sequence:
 \begin{equation*}
     \begin{tikzcd}
    0\arrow{r} & \mathcal{A}_{[1]}\arrow{r}{i} & \mathbb{G}\arrow{r}{\pi} & G\arrow{r}\arrow[bend left=33]{l}{\varphi} & 0 \, , \\
\end{tikzcd}
 \end{equation*}
 where we arbitrarily choose a splitting $\varphi$ at the level of groupoids.  The complete data of $\varphi$ is that for all $g \in G$, $\varphi (g)$ is an object in $\mathbb{G}$ and thus $\mathbb{G} \cong \mathcal{A}_{[1]} \rtimes_\varphi G $. Such $\mathbb{G}$ is a groupoid with group law and $(\ast, g)\in \ob(\mathbb{G})$; this is known as a 2-group.  The associator $\beta_{\mathbb{G}}$ in $\mathbb{G}$ is a 3-cocyle valued in the group $A$ and gives the isomorphism
\be
m_{\mathbb{G}}[m_{\mathbb{G}}[(\ast, g_1), (\ast, g_2)],(\ast, g_3))] \simeq m_{\mathbb{G}}[(\ast, g_1),m_{\mathbb{G}}[(\ast, g_2),(\ast, g_3)]] 
\ee
 where $m_\mathbb{G}$ is the multiplication of objects in the groupoid.  A different choice $\varphi'$ of splitting which is isomorphic to $\varphi$ changes $\beta_{\mathbb{G}}$ by an exact term.  Therefore, isomorphism classes of $\varphi$ determine $\beta_{\mathbb{G}}$ as a class in the cohomology $\rH^3(G;A)$.   In the literature, this class is known as the Postnikov class and is said to take values in the one-form symmetry, viewed in its full form as a groupoid.  This is convenient because it allows us to view the associator as a topological defect that implements the one-form symmetry; in (1+1)$d$ this is a codimension two operator.

 We will come back to how this one form symmetry defect can be related to the lack of associativity of composing zero-form symmetry defects in a later subsection.    We furthermore point out that, as can be seen from the above short exact sequence, $\mathcal{A}_{[1]}$ is a subgroupoid of $\mathbb{G}$, but $G$ need not be.  Therefore, one could always ask the question of gauging the one-form symmetry in a 2-group, but not necessarily the zero-form. 
 
 A more topological viewpoint of 2-groups is from the point of view of its classifying space. The classifying space of a two group is a $\textit{homotopy 2-type}$ \footnote{This is only true for geometrically discrete 2-groups that we are considering, and not generally for 2-group analogues of Lie groups.} \cite{Debray:2018}, meaning it only has two nontrivial homotopy groups.  Consider a connected space $\rB \mathbb G$  such that: 
 \begin{equation}
     \pi_1 (\rB \mathbb G) \cong G, \quad \pi_2(\rB \mathbb G) \cong A.
 \end{equation}
 In order to have the two symmetries $G$ and $A$ mix, take $\rB \mathbb G$ as a fiber bundle over $\rB G$ with fiber $K(A,2)$. We can specify the bundle $p: \rB \mathbb G \to \rB G$ by its homotopy cofiber i.e., a map $k: \rB G \to \Sigma K(A,2) \cong K(A,3)$ known as the \textit{k-invariant} of the space $\rB \mathbb{G}$.  The associator $\beta_\mathbb G$ was a class in the third cohomology, which can be represented by maps into Eilenberg-Mac Lane spaces. This is to say $\rH^3(G; \cA_{[1]}) \cong [K(G,1), K(A,3)]$, which sends the associator of $\mathbb G$ to the k-invariant of $\rB \mathbb{G}$. Moreover, every homotopy 2-type $\rB \mathbb{G}$ is the classifying space of some 2-group $\mathbb G$.  
 
 \subsection{Symmetry Defects and 2-group Structure}\label{SymDefects}
 We now study a physical system with 2-group symmetry at the level of its symmetry defects.  We choose a triangulation of the space with a defect network of lines corresponding to coupling the zero-form symmetry to background connection \cite{Tachikawa:2019}.   The line defects in a (1+1)$d$ theory form a 1-monoidal 1-category where the trivalent junction of line defects is codimension two and serve as morphisms between codimension one objects.  In general, a $p$-monoidal $q$-category contains objects of dimension $q$ for which there are $p$ ambient dimensions to compose object. Another way to understand this type of category is to envision it as a way of encapsulating the ways to  ``compose" $(q-1)$-branes in $(p+q)$-dimensions of spacetime.  
 
 In an $n$-dimensional theory, the natural object to study is an``$n$-group" in which one considers the topological operators that implement $k$-form symmetries for $k \leq (n-1)$. In analogy with the case in (1+1)$d$, the objects of codimension one have morphisms that are codimension two, and the codimension two objects have morphisms that are codimension three, etc.  This results in the notion of a weak $n$-category, where the compositions are only associative up to higher coherence relations.  In physical systems of interest we must also implement the condition that the $n$-category has trivial center in the sense that no operator can ``commute'' in a higher categorical sense with all other operators in the theory.  If there is a nontrivial center in the category, then we deem that there is a gravitational anomaly and therefore obstructs the consistent realization of the category as a physical system purely in $n$-dimensions \cite{Kong:2014qka}.
 
 In (1+1)$d$ the zero-form symmetry operators of $\textbf{g},\textbf{h},\textbf{k}$ can be composed in two ways, and going between the two is a matter of applying an $F$-move.  As depicted in figure \ref{F-move Postnikov}, a point operator $\beta_{\mathbb{G}}(\textbf{g},\textbf{h},\textbf{k})$ valued in the one-form symmetries can be created due to this move, signifying a nontrivial interaction between the two types of symmetries and therefore a two group structure.  The $F$-move involving the zero-form symmetry defect could furthermore generate a phase $\omega(\textbf{g},\textbf{h},\textbf{k}) \in \rC^3(G, \rU(1))$ that is purely attributed to the fact that the Hilbert space is in a projective representation of the zero-form symmetry.  We will later explicity calculate anomalies of this type for full 2-group theories using techniques in group cohomology.   
 
\begin{figure}
    \centering
    \begin{tikzpicture}[thick, scale = .8]
    \draw[line width = .5mm] (4,4) -- (1,1);
    \draw[line width = .5mm] (-2,4) -- (1,1);
    \draw[line width = .5mm] (1,1) -- (1,-2);
    \draw[line width = .5mm] (2.5,2.5) -- (1,4);
    \draw[right] (-2.5,4.5) node {$\textbf{g}$};
    \draw[right] (.5,4.5) node {$\textbf{h}$};
     \draw[right] (4,4.5) node {$\textbf{k}$};
     \draw[right] (.5,-2.4) node {$\textbf{ghk}$};
     \draw[right] (3.9,.5,1) node {$\omega(\textbf{g},\textbf{h},\textbf{k})$};

    \draw[line width = .5mm] (4+8,4) -- (1+8,1);
    \draw[line width = .5mm] (-2+8,4) -- (1+8,1);
    \draw[line width = .5mm] (1+8,1) -- (1+8,-2);
    \draw[line width = .5mm] (2.5+8-3,2.5) -- (1+8,4);
    
    \draw[right] (12,4.5) node {$\textbf{k}$};
    \draw[right] (5.6,4.5) node {$\textbf{g}$};
     \draw[right] (9,4.5) node {$\textbf{h}$};
     \draw[right] (8.5,-2.4) node {$\textbf{ghk}$};

    \draw[->] (4.5,0,1) -- (6,0,1);
    
    \tikzstyle{vertex}=[circle,fill=black!100, minimum size=1pt,inner sep=3pt]
    \node[vertex] at (9,1){};
    \draw[right]  (6.3,.5) node {$\beta_{\mathbb{G}}(\textbf{g},\textbf{h},\textbf{k})$};
    \end{tikzpicture}
        \caption{}
    \label{F-move Postnikov}
\end{figure}

Another feature in (1+1)-dimensions is that point operators, which are charged under the zero-form global symmetry and pass through a symmetry line defect, are acted upon by the line, see figure \ref{remote detect}.  Therefore, the point operator labeled by $\textbf{a}$ becomes $\textbf{a}_\textbf{g}$ after passing through a $\textbf{g}$-defect.   If $\textbf{a}$ was the only unique point operator that existed in a theory, then $\textbf{a}_{\textbf{g}} = \textbf{a}$ which implies that the nontrivial $\textbf{g}$-defect could not even be detected\footnote{The fact that all operators need to be detectable goes by the name \textit{remote detectability}.  In topological orders such a condition is necessary in order to have a consistent anomaly free theory. }. This means that in order to consider nontrivial zero-form operators, we must have at least two independent point operators. In theories living in (2+1)-dimensions, a single unique point operator would suffice because lines can detect other lines through their braiding, and lines can detect surfaces by puncturing at points, see figure \ref{linedown}. 

\begin{figure}[ht]
\centering
    \begin{tikzpicture}[thick, scale = .8]
        \def\Depth{4}
        \def\DepthTwo{3}
        \def\Height{2}
        \def\Width{2}
        \def\Sep{3}
    
    \tikzstyle{vertex}=[circle,fill=black!80, minimum size=8pt,inner sep=2pt]
    \draw[left]  (-1.2-4,.25) node {$\textbf{a}$};
    \node[vertex]at ( -1-4 , 0) {};
    \draw[black, line width = .5mm] (.5-4,-2) -- (.5-4,2) ;
    \draw[black] (.5-4,2.5) node{$\textbf{g}$};
     \draw[->,decorate,decoration={snake,amplitude=.4mm,segment length=2mm,post length=1mm}] (\Height-1-3,0) -- (\Sep-1-3, 0);
     \draw[black, line width = .5mm] (3.5-3,-2) -- (3.5-3,2) ;
     \draw[black] (.5,2.5) node{$\textbf{g}$};
      \node[vertex]at ( 2 , 0) {};
     \tikzstyle{vertex}=[circle,fill=black!80, minimum size=4pt,inner sep=1pt]
     \draw[right]  (2.2,.25) node {$\textbf{a}_\textbf{g}$};
	\end{tikzpicture}
	\caption{}
	\label{remote detect}
\end{figure}
The local point operators determine the ground states of our system, which means that theories in (1+1)$d$ are most interesting to study in the presence of multiple ground states, or vacua.  For the remainder of this note we will use the terms ``ground state" and ``vacua" interchangeably.  Going back again to (2+1)-dimensions, one could also consider multiple ground states, and in general there will exist a modular tensor category (MTC) describing the information and interactions of anyons in the theory around each ground states.  The theory with respect to a single ground state is what is usually referred to in the traditional discussion of topological order, in which case it is described by a \textit{fusion n-category}.   We see that in order to give a more complete treatment of topological orders, and general theories in (1+1)-dimensions, one needs to modify the definition to take into account the possibility of multiple ground states and a decomposition into subsectors.  A more complete definition of topological orders in $(n+1)$-dimensions with multiple ground states is as a \textit{multifusion n-category} \cite{Johnson-Freyd:2020usu}. 

Suppose we are presented with a theory $\mathcal{T}$ in (1+1)-dimensions with some global zero-form symmetry group $G$, and that $\mathcal{T}$ has multiple vacua when considered at some finite energy.   If the system is at any particular vacua, then at finite energy there could be instantons that tunnel between different vacua.   We therefore do not say that these vacua can be regarded independently, but rather should be thought of as a family of vacua.  By flowing to the deep infrared (IR), all the massive instantons are integrated out, and it is sensible to claim that the system indeed sits at a particular vacuum. Due to the potential between the vacua becoming arbitrarily high in the IR, we regard each point operator as also labeling a subsector of our theory where each subsector is decoupled and truly independent of the others.

This vacuum at which our system sits is labeled by a point operator in the set of $\mathcal{S}=\{ \widehat1,\,\widehat2,\ldots,\, \widehat M\}$ where $M$ is the cardinality of $\widehat A = \hom(A,\rU(1))$, the Pontrjagin dual group of $A$.  We see that the one form symmetry is an emergent symmetry of $\mathcal{T}$ in the deep IR. The algebra of point operators is a finite dimensional commutative and separable algebra and is isomorphic to a direct sum indexed by $\spec(\mathcal{A}_{[1]})(\bC)$.  Thus, we can write $\mathcal{T}$ in the deep IR as a direct sum of its subsectors $\mathcal{T} = {\mathcal{T}_{\widehat1} \oplus \mathcal{T}_{\widehat2} \oplus \ldots \oplus \mathcal{T}_{\widehat M}}$.   The theories we will be most interested in for this paper are those that exhibit the properties of $\mathcal{T}$ in the far IR. We note in passing that in (0+1)-dimensions, a zero-form symmetry $G$ alone will split theory $\mathcal{Q} = \bigoplus_{\hat\alpha \in \widehat G}  \mathcal{Q}_{\hat\alpha}$, where $\widehat{G}$ is the Pontryagin dual to $G$.  

In a local patch of the theory with two insertions of point operators, we could imagine composing the two points and producing another point operator, therefore endowing this set of operators with a multiplication structure.  To better illustrate composing one-form operators and the relation with degenerate ground states, we give an example again in the (2+1)$d$ world but consider one of the spatial dimensions compactified on a circle.  Along this compactified direction we may wrap a number of anyons of which some may generate a one-form symmetry.  Each wrapping of an anyon labels a ground state. We may decide to bring two anyons labeled by $a$ and $b$ close together and fuse them using the rule $ a \times b = \sum_c \,N^{c}_{ab}\,c $.  For an abelian anyon $a$, this fusion gives just a single anyon; we see that the multiple vacua give a representation of the fusion of anyons.

In a general $(n+1)$-dimensional theory, the $n$-dimensional operators do not only have a fusion, or monoidal $n$-category structure, but it is also possible to produce composite operators by taking direct sums. Therefore, it is also possible to consider the situation in which a complex linear combination of multiple operators in the set $\mathcal{S}$ as solely operators that do not interact with each other in the way of fusion, giving the addition structure.  We denote $\mathbb{C}[\mathcal{A}_{[1]}]^\times$ as the ring of invertible one form operators with the ring structure as previously laid out.

\subsubsection{Examples of multiple ground states}

To get a better handle on how to understand multiple ground states and the subtleties which arise from the decomposition into subsectors we consider some examples.  Let us consider a (2+1)$d$ topological quantum field theory (TQFT) described by a MTC $\mathcal{C}$, such that the category $\mathcal{C}$ has as its boundary a rational conformal field theory (RCFT) that consists of a pair of unitary vertex operator algebras (VOA) $V,W$ and an equivalence of categories $\Phi : \operatorname{Rep}(V) \overset\sim\to \operatorname{Rep}(W)^{\mathrm{op}}$.  The (1+1)$d$ operators that arise from a (2+1)$d$ object can be lines, obtained by running an operator of codimension 1 parallel to the top and bottom of the slab, or local point operators by letting a line in bulk end on the sides of the slab, as in figure \ref{BulkBoundary}.  In particular, an object in $\cC$ contains a vector space of ways to end, which is in fact a module for the VOAs. 
\begin{figure}[ht]
\centering
    \begin{tikzpicture}[thick,scale=.7]
        \def\Depth{7}
        \def\Height{2}
        \def\Width{2}
        \coordinate (O) at (0,0-1,0-1);
        \coordinate (A) at (0,\Width+1,0-1);
        \coordinate (B) at (0,\Width+1,\Height+1);
        \coordinate (C) at (0,0-1,\Height+1);
        \coordinate (D) at (\Depth,0-1,0-1);
        \coordinate (E) at (\Depth,\Width+1,0-1);
        \coordinate (F) at (\Depth,\Width+1,\Height+1);
        \coordinate (G) at (\Depth,0-1,\Height+1);
        
        \draw[above] (\Depth/2,\Width+1/2+1.5,\Height/2) node {\large $\mathcal{T}\times[0,1]$};
        \draw[left] (-1, \Width/2, \Height/2) node {\large $V$};
        \draw[right] (\Depth+1, \Width/2, \Height/2) node{\large $ {W}$};
        
        \draw[black] (O) -- (C) -- (G) -- (D) -- cycle;
        \draw[black] (O) -- (A) -- (E) -- (D) -- cycle;
        \draw[black,fill=green!10] (O) -- (A) -- (B) -- (C) -- cycle;
          \draw[black,fill=black!10,opacity=.7] (0,1.3,\Height+1) -- (\Depth,1.3,\Height+1)--(\Depth,1.3,-1)--(0,1.3,-1)--cycle;
        \draw[black,fill=red!20,opacity=.8] (D) -- (E) -- (F) -- (G) -- cycle;
        \draw[black] (C) -- (B) -- (F) -- (G) -- cycle;
        \draw[black] (A) -- (B) -- (F) -- (E) -- cycle;

        \draw[decoration={markings, mark=at position 0.5 with 
 {\arrow{>}}}, postaction={decorate}] (0,\Width/5,\Height/2) .. controls (0.3*\Depth+1,-1.1*\Width+2.5,1.3*\Height)   .. (\Depth,\Width/5,\Height/2);
 
 \draw[decoration={markings, mark=at position 0.5 with 
 {\arrow{>}}}, postaction={decorate},opacity=.35] (0,1.3,\Height+1) -- (0,1.3,-1);
 
  \draw[decoration={markings, mark=at position 0.5 with 
 {\arrow{>}}}, postaction={decorate},opacity=.5] (\Depth,1.3,\Height+1) -- (\Depth,1.3,-1);
 

 \tikzstyle{vertex}=[circle,fill=black!100, minimum size=.01pt,inner sep=1pt]
    \node[vertex] at (0,\Width/5,\Height/2) {};
    
     \tikzstyle{vertex}=[circle,fill=black!100, minimum size=.01pt,inner sep=1pt]
    \node[vertex] at (\Depth,\Width/5,\Height/2) {};
 
    \end{tikzpicture}
    \caption{  }
    \label{BulkBoundary}
\end{figure}
For $X\in \mathcal{C}$ traversing between the left and right face, let $V(X)$ be the corresponding $V$-module, and $W(X)$ for the corresponding $W$-module.  The full algebra of local operators is:
\begin{equation}
   \mathcal{A}  = \bigoplus_{X \in I} V(X) \otimes W(X)
\end{equation}
where $I$ is a set of representatives of simple objects.  Each $V(X) \otimes W(X)$ has an action of the left and right Virasoro algebra $\operatorname{Vir}_L \otimes \operatorname{Vir}_R$, and so of $L_0,\, \bar{L}_0$; the only state with $L_0 = \bar{L}_0 = 0$ is $ \ket{0} \otimes \overline{\ket{0}}$ which is the local ground state, in this case also the vacuum. 

This  axiomatization describes  the  local  operators of the CFT, but we note that there was no mention of a  Hilbert  space of states on the interval. Due to the lack of that extra information, automorphisms of an RCFT $(V,W,\Phi)$ can have ’t Hooft anomalies, and one must consider the possibility of other ``anomaly” type data.  If we have an RCFT with two local ground states we can take their direct sum, then we should have data $(V_1,W_1,\Phi_1)$ for the first ground state, and $(V_2,W_2,\Phi_2)$ for the second.  We need some extra information which comes in the form of some anomaly trivializing information, consisting of a choice of ``category of walls" between the two ground states.
Looking at the full operator content, the two vacua turn the data  into a $2\times 2$ block matrix:
\begin{equation}
\begin{pmatrix}
    \begin{array}{c|c}
    (V_1,W_1,\Phi_1) & \text{extra data}\\ \hline
     (\text{extra data})^{T} & (V_2,W_2,\Phi_2)
    \end{array}
    \end{pmatrix}
\end{equation}
where the ``extra data" encodes the passage from one ground state to the other.  Another way to view this direct sum is to consider instead of working purely in 2$d$, working with a pair consisting of 
\begin{equation*}
    \{\text{(1+1)}d \,\, \text{boundary condition},\,\text{(2+1)}d \,\,\text{theory} \}.
\end{equation*}
  The operator content as well as the anomalous data of the 2$d$ theory determines an absolute 2$d$-3$d$ theory i.e. the data $(V_1,W_1,\Phi_1)$ leads to data about a cobordism invariant of 3-manifolds coming from the central charges of $(V_1,W_1,\Phi_1)$ where the operators in 3$d$ are the center of the operators in 2$d$.  We can take direct sums of absolute 2$d$-3$d$ theories, by summing in both 2$d$ and 3$d$. 
  
  A concrete realization of the extra data that can appear between two ground states can be observed in anyon condensation. This is gives a way of interpolating between going between two different (2+1)$d$ topological orders by deforming adiabatically.  In the categorical language we start off with a MTC $\cC$ and consider some condensible algebra $\alpha$, built by condensing an abelian line $a \in \cC$ of integer spin that generates the one-form symmetry with the  vaccuum \cite{Hsin:2018vcg, Kong_2014}.  Performing the condensation involves projecting out lines that have nontrivial monodromy charge with $a$.  We then land in a new MTC $\cC'$ separated from $\cC$ by a gapped domain wall.  The domain wall excitations are given by the category of $\alpha$-modules in $\cC$ which can fuse but not braid with each other.   
  
  Let us furthermore consider the case in which a $G$-symmetry acts on the local ground states. In special cases, the $G$-symmetry may protect a degeneracy in ground states, by forbidding any small deformation in the form of local operators that can distinguish between ground states.  By turning off the symmetry we might expect a decomposition into a direct sum of subtheories, but more information is necessary.  Namely, the data regarding the relative phases between the ground states. At a particular ground state, it is possible to stack with a trivial, or invertible theory. Such theories are invertible from the point of view of field theory, which says we have a map 
  $\alpha:$ \textbf{Bord}$_3 \to$ \textbf{Vec}$_\bC$ from three dimensional bordisms of two manifolds to specifically the subset of \textit{invertible vector spaces}.  Invertibility of vector spaces is given by invertibility under the tensor product, and a vector space is only invertible under tensor product if it is one dimensional. Hence, we require $\alpha(Y^{2}) \in$   \textbf{Vec}$_\bC$ to be a line.  Invertible field theories are conjectured to describe the low energy limits of symmetry protect topological (SPT) phases, where a symmetry $G$ protects the phase from being gapless.   We will associate the language of ``stacking with SPT" in order to refer to the relative information that can exist between ground states.  
  %
  In (1+1)$d$, these SPTs are classified by $\rH^2(G\,;\rU(1))$.  If however, as in many cases, $G$ has an action on the coefficients, then this \textit{twisted cohomology} may be nontrivial. More precisely, for any module $\rho: \bZ[G] \to \operatorname{Aut}(A)$ over $G = C_N = \langle t\rangle$ cyclic, we have:
  \begin{equation}
    \rH^\bullet(G; A) = \begin{cases}
      A^G \text{ fixed points}, & \bullet = 0, \\
      \ker(\rho(N)) / \im(\rho(t)-1), & \bullet = \text{odd}, \\
      \ker(\rho(t)-1) / \im(\rho(N)), & \bullet = \text{even} > 0.
    \end{cases}
\end{equation}
Here $N = 1+ t +t^2 +\ldots + t^{n-1}$.  As an explicit example, take $G = \bZ_2^T$ and $A = \rU(1)$, where $\bZ_2^T$ acts by complex conjugation on $\rU(1)$.  Then $\rH^2(\bZ_2^T\,;\rU(1))= \bZ_2$, which means we have two choices of SPT that can manifest as information of a relative phases between the ground states.  

\subsection{2-group Background Gauge Fields}
The decomposition structure of theories in (1+1)$d$ implies that we must modify our 2-group ingredients to take this into account.   
  The Postnikov $\beta_{\mathbb{G}} $, which is a class in $\rH^3(G\,; \cA_{[1]})$, must now be modified to encompass the statement that the lack of associativity in $G$-symmetry defects can manifest as a complex linear combination of invertible one-form operators at a trivalent junction of $G$-defects.  There exists a group homomorphism $f: \cA_{[1]} \to \mathbb{C}[\cA_{[1]}]^\times$ that acts functorially on cohomology.  Therefore, we can build $[f(\beta_{\mathbb{G}})] \in \rH^3(G\,;\,\mathbb{C}[\cA_{[1]}]^\times)$.
  We now use the fact that there is a ring isomorphism of 
\begin{equation}
\mathbb{C}[\cA_{[1]}]^\times = \bigoplus_{\widehat A} \mathbb{C}^\times \,,   
\end{equation}
to express $[f(\beta_{\mathbb{G}})]$ as a class 
\begin{equation}\label{FofBeta}
     [f(\beta_\mathbb{G})] \in \bigoplus_{\widehat A} \rH^3(G\,; \bC^\times) = \bigoplus_{\widehat A} \rH^3(G\,; \rU(1)).
\end{equation}
We will denote a 2-group theory $\mathcal{T}$ with partition function $\mathcal{Z}^{\mathcal{T}}_{A^{(1)};\,B^{(2)}}$, where $A^{(1)}$ and $B^{(2)}$ are the background gauge fields that couple to the zero-form symmetry, and one-form symmetry respectively.  The upper indices denote the fact that an $n$-form symmetry couples to an ($n$+1)-form background gauge field. Under gauge transformations the fields transform as \cite{Benini:2018reh,Sati_2009,fiorenza2011cech}
\begin{equation}\label{2groupTransform}
    A^{(1)} \to A^{(1)} + \frac{1}{N}d\lambda^{(0)} ,\hspace{2mm} B^{(2)} \to B^{(2)} + \frac{1}{M}d \Lambda^{(1)}+  \frac{{\beta_{\mathbb{G}}}}{N M}\lambda^{(0)} \cup d A^{(1)}\,,
\end{equation}
where $\lambda^{(0)}$ and $\Lambda^{(1)}$ are zero and one cochains for the background $G$ and $\cA_{[1]}$ symmetries respectively. The transformation in $B^{(2)}$ involving a mixture between the two symmetries is characteristic of a 2-group theory, and is parametrized by the Postnikov.  The cup product is used in the way which is described in \cite{Fiorenza_2013}, for cup
product Chern-Simons theories.

While the Postnikov was touted as a topological defect in \S\ref{SymDefects} valued in groupoids, here in the gauge transformations $\beta_{\mathbb{G}}$ takes a numerical value as an integer modulo $M$ representing the class $[\beta_{\mathbb G}] \in \rH^3(G\,; {A})$\footnote{It is important to keep track of what category $\beta_{\mathbb{G}}$ is valued in so we can distinguish between (mixed) anomalies and 2-groups.}.  For convenience, from now on we normalize all the background connections for discrete symmetry groups so that there is no need to divide by the order of the group.  Any integral over a gauge field is understood to be analogous to a discrete Fourier transformation.
The partition function $\mathcal{Z}^{\mathcal{T}}$ attains a term under gauge transformation which takes the form 
\begin{equation}
    \exp{\left(i\,\beta_{\mathbb{G}}\, \int \lambda^{(0)}\cup dA^{(1)} \right) }.
\end{equation}
But, because the Postnikov class should now more precisely be thought of as the class $[f(\beta_{\mathbb{G}})]$, the gauge field $B^{(2)}$ in principle could have different 2-group gauge transformation on each sector of the original theory
\begin{equation}
   A^{(1)}_{\widehat\alpha} \to A^{(1)}_{\widehat\alpha} + d\lambda^{(0)}_{\widehat\alpha} ,\hspace{2mm} B^{(2)}_{\widehat\alpha}  \to B^{(2)}_{\widehat\alpha} + d \Lambda^{(1)}_{\widehat\alpha}+ f(\beta_{\mathbb{G}})_{\widehat\alpha} \,\lambda^{(0)}_{\widehat\alpha}\cup dA^{(1)}_{\widehat\alpha},
\end{equation}
and therefore the partition function also attains a term on each sector of the form
\begin{equation}\label{2group}
     \bigoplus_{\widehat{\alpha}}\CZ^{\mathcal{T}}_{A^{(1)}_{\widehat\alpha};\,B^{(2)}_{\widehat\alpha}}\,\exp\left(i\,f(\beta_{\mathbb{G}})_{\widehat\alpha} \,\int  \lambda^{(0)}_{\widehat\alpha} \cup dA^{(1)}_{\widehat\alpha}\right).
\end{equation}
We remark that even if $f(\beta_{\mathbb{G}})_{\widehat\alpha} $ does not vanish, this phase should not be considered an anomaly in the usual sense of 't Hooft anomalies. Since we are strictly speaking dealing with a zero-form symmetries extended by one-form symmetries, the change in the partition function under gauge transformation is controlled by the extension, which manifests as a one form operator.   

\subsection{Gauging in a 2-group theory}
Having established the formalism for the partition function and gauge transformations, in this section we begin to manipulate the theory at the level of its partition function.
For $\mathcal{T}$ in which $\mathcal{A}_{[1]}$ acts nonanomalously, we can  ask to gauge this symmetry. After this we land on a theory denoted  $\mathcal{T}/\!/ \mathcal{A}_{[1]}$.  We will see that upon gauging the one-form symmetry, $\mathcal{T}/\!/ \mathcal{A}_{[1]}$ will have a mixed anomaly between the zero-form symmetry and the ``dual" symmetry, which is a $(-1)$-form symmetry, $ \widehat{\cA}_{[-1]}$, controlled by the Postnikov, from the ungauged 2-group.  To tell this story in a more familiar way, we outline the gauging procedure with zero-form symmetries; the logic carries over to one-form symmetries upon shifting some indices.  
Suppose we have a theory $\mathcal{T}'$ in $d$-dimensions with an action of $ \tilde{G} = G_{[0]} \rtimes_\beta A_{[0]}$ where $\beta \in \rH^2(G_{[0]};A_{[0]})$, so that $\tilde{G}$ is an extension by two zero-form symmetries. In gauging $A_{[0]}$ in $d$-dimensions we expect to get a dual group that is a $\widehat{A}_{[d-2]}$ in  $\mathcal{T'}/\!/{A}_{[0]}$ and there could be a purely mixed anomaly between $ G_{[0]} \times \widehat{A}_{[d-2]} = \widehat{G}$ which is a class in $\rH^{d+1}(\widehat{G}\,;\rU(1))$.  This cohomology can be calculated by a spectral sequence, in which on the $E_2$ page we have  $\rH^p(G;\rH^q(\widehat{A}_{[d-2]}\,;\rU(1)))$, that converges to $\rH^{p+q}(\widehat{G};\rU(1))$.  We have that $\rH^{d-1}(\widehat{A}_{[d-2]}\,;\rU(1)) = \Hom(\widehat{A}_{[d-2]}\,,\rU(1)) = A_{[0]}$, and $\rH^2(G;\rH^{d-1}(\widehat{A}_{[d-2]}\,;\rU(1)))$ converges to $\rH^{d+1}(\widehat{G};\rU(1))$.  This is interpreted to mean that the mixed anomaly is given by the extension $\beta$ from the original $\mathcal{T}'$ theory, by cupping with a $(d-1)$-cochain valued in $\widehat{A}_{[d-2]}$.

One can also tell this story in reverse. Suppose that a theory $\tilde{\mathcal{T}}'$ is acted upon by the group\, $G \times \widehat{A}_{[d-2]}$ and there exists a purely mixed anomaly $\alpha \in \rH^2(G\,;\rH^{d-1}(\widehat{A}_{[d-2]}\,;\rU(1)))$.   In gauging $\widehat{A}_{[d-2]}$, we expect a ${A}_{[0]}$  symmetry in $\tilde{\mathcal{T}}' /\!/ \widehat{A}_{[d-2]}$.  The analogue of  2-groups involving $G$ and ${A}_{[0]}$ is controlled by an extension in  $\rH^2(G\,;{A}_{[0]}) = \\
\rH^2(G\,;\rH^{d-1}(\widehat{A}_{[d-2]}\,;\rU(1)))$, which is $\alpha$.   We see through this spectral sequence argument that a theory with a mixed anomaly, when gauged, becomes a theory with 2-group symmetry, where the class of the anomaly becomes the extension defining the 2-group.  

When we focus specifically for our theory $\mathcal{T}$ in (1+1)-dimensions, then our extension is valued in $\H^3(G\,;\mathbb{C}[\cA_{[1]}]^\times)$.  Due to the functorality of the spectral sequence, there is a homomorphism from the spectral sequence of $\mathbb{C}[\cA_{[1]}]^\times.\,G$ to the spectral sequence of $\cA_{[1]}.G$, here  as $X.Y$ denotes the extension of $Y$ by $X$.  The homomorphism is given on the $E_2$ page by
\begin{equation}
 \rH^\bullet(G;f^\bullet) : \rH^\bullet(G; \rH^\bullet(\mathbb{C}[\cA_{[1]}]^\times;\rU(1))) \to  \rH^\bullet(G; \rH^\bullet(\cA_{[1]};\rU(1)))\,,
 \end{equation}
 where
\begin{equation}
    f^\bullet : \rH^\bullet(\mathbb{C}[\cA_{[1]}]^\times; \rU(1)) \to \rH^\bullet(\cA_{[1]};\rU(1)) 
\end{equation}
is the pull back of the homomorphism $f: \cA_{[1]} \to \mathbb{C}[\cA_{[1]}]^\times$.  This means that upon gauging, it makes sense to take the restriction of $f(\beta_\mathbb{G})$ to each element $\widehat{\alpha}\in \widehat{\cA}_{[1]}$, and therefore there is a notion by which the mixed anomaly can be restricted over any particular subsector into which a theory decomposes. 

We now perform the gauging explicitly at the level of partition function.  In the following, we use lower case letters to denote that the background gauge field is being integrated over in the path integral. In order to gauge the one form symmetry in $\mathcal{Z}^{\mathcal{T}}$ we integrate over the two-form gauge field \cite{Cordova:2018cvg}
\begin{equation}\label{GaugeOneForm}
    \mathcal{Z}^{\mathcal{T}/\!/ \mathcal{A}_{[1]}}_{A^{(1)};\,C^{(0)}}=\int D b^{(2)}\, \mathcal{Z}^{\mathcal{T}}_{A^{(1)};\,b^{(2)}}\exp{\left(-i  \,\int C^{(0)} \cup b^{(2)} \right)},
\end{equation}
where $C^{(0)}$ is the background gauge field for the dual $(-1)$-form symmetry \cite{Cordova:2019jnf} in $\mathcal{T}/\!/ \mathcal{A}_{[1]}$.  One can furthermore gauge the $(-1)$-form symmetry by summing over the zero-form gauge field.  This takes us back to $\mathcal{T}$:
\begin{align}
    &\int D c^{(0)}\,\mathcal{Z}^{\mathcal{T}/\!/ \mathcal{A}_{[1]}}_{A^{(1)};\,c^{(0)}}\exp{\left( i \,\int B^{(2)}\cup c^{(0)}\right)}\notag\\[2pt]
    &= \int D c^{(0)} D b^{(2)}\,\mathcal{Z}^{\mathcal{T}}_{A^{(1)};\, b^{(2)}}\exp{\left(-i \,\int c^{(0)}\cup b^{(2)}\right)}\exp{\left( i \, \int B^{(2)}\cup c^{(0)}\right)}\notag\\[2pt]
    & = \mathcal{Z}^{\mathcal{T}}_{A^{(1)};\,B^{(2)}}.
\end{align}
We now move on to the case of a two group with $\beta_\mathbb{G}$, and again gauge the one form symmetry in the manner of \eqref{GaugeOneForm}.  If we implement the 2-group transformation in \eqref{2groupTransform} we get 
\begin{align}\label{GaugedPartition}
    \mathcal{Z}^{\mathcal{T}/\!/ \mathcal{A}_{[1]}}_{A^{(1)};\,C^{(0)}} &=\int D b^{(2)}\, \mathcal{Z}^{\mathcal{T}}_{A^{(1)};\,b^{(2)}} \exp\left(-i\, \int C^{(0)} \cup b^{(2)} \right) \notag\\
    &\quad \times \exp \left( -i \, \int C^{(0)} \cup d\Lambda^{(1)}\right) \exp{\left(-i\,  \int  C^{(0)} \cup \left(  \, {\beta}_{\mathbb{G}}  \cup \lambda^{(0)} \cup \d A^{(1)}\right)  \right)}\,, \notag \\
\end{align}
where $ C^{(0)}$ and $\lambda ^{(0)} \cup dA^{(1)} $ take value in  $\widehat{\cA}_{[-1]}$ and $G$ and the Postnikov we take to be valued in $\cA_{[1]}$ instead of its c-number value as in \eqref{2groupTransform}. We claim that the $(-1)$-form symmetry is best thought of as $\spec(\mathcal{A}_{[1]})(\mathbb{C}) = \hom(\mathcal{A}_{[1]}, \mathbb{C})$, and presently justify this proposition.   Inside the exponential, cupping $C^{(0)}$ with $\beta_\mathbb{G}$ is this map and gives a c-number contribution multiplying the curvature of the background $G$ gauge field.  If we consider attaching a term in the above partition function of the form 
\begin{equation}\label{cupwithC}
    \exp{\left(i\,\left( C^{(0)} \cup \beta_{\mathbb{G}}\right)\, \int \lambda^{(0)}\cup \d A^{(1)}\right) },
\end{equation}
this will exactly cancel out the exponential on the right hand side in  \eqref{GaugedPartition}.  This factor is a mixed anomaly in the gauged theory, which is controlled by the extension $\beta_{\mathbb{G}}$ we started off with in the 2-group ungauged theory.
Going in reverse, consider a mixed anomaly classified by $\widehat{\beta}_{\tilde{\mathbb{G}}} \in \rH^3(G\,,\widehat{\cA}_{[-1]})$ in the theory $\mathcal{T}/\!/\mathcal{A}_{[1]}$ and gauge the $\widehat{\mathcal{\cA}}_{[-1]}$ symmetry by integrating over $C^{(0)}$. 
 The partition function takes the form 
\begin{align}
     \cZ&^{\mathcal{T}/\!/\mathcal{A}_{[1]}}_{A^{(1)};\,C^{(0)}}\,\exp{\left(i \widehat{\beta}_{\tilde{\mathbb{G}}} \, \int \lambda^{(0)}\cup \d A^{(1)}\cup C^{(0)} \right)}\notag\\[2pt]
     &\overset{\text{gauge}}\longrightarrow \int D c^{(0)} \cZ^{(\mathcal{T}/\!/\mathcal{A}_{[1]})/\!/\widehat{\mathcal{A}}_{[-1]}}_{A^{(1)};\,c^{(0)}}\exp{\left(i\widehat{\beta}_{\tilde{\mathbb{G}}} \,\int  \lambda^{(0)}\cup \d A^{(1)}\cup c^{(0)}\right)}\notag\\[2pt]
     &\hspace{60mm}\times \exp{\left(i\, \int B^{(2)}\cup c^{(0)}\right)}\notag\\[2pt]
     &= \int D c^{(0)} Db^{(2)}\cZ^{\mathcal{T}}_{A^{(1)};\,b^{(2)}}\exp{\left(i\widehat{\beta}_{\tilde{\mathbb{G}}} \, \int \lambda^{(0)}\cup \d A^{(1)}\cup c^{(0)}\right)}\notag\\[2pt]
     &\hspace{60mm}\times \exp{\left(-i\,\int  c^{(0)}\cup b^{(2)}\right)}\exp{\left(i\, \int B^{(2)}\cup c^{(0)}\right)}\notag\\[2pt]
     & = \cZ^{\mathcal{T}}_{A^{(1)};\, B^{(2)}+\widehat{\beta}_{\tilde{ \mathbb{G}}} \,\int\lambda^{(0)}\cup dA^{(1)}}\notag\\[2pt]
     &= \cZ^{\mathcal{T}}_{A^{(1)};\,B^{(2)}}\exp{\left(i\widehat{\beta}_{\tilde{\mathbb{G}}} \,  \int \lambda^{(0)}\cup \d A^{(1)}\right)},
\end{align}
which is exactly the transformation in a two group theory with Postnikov $\widehat{\beta}_{\tilde{\mathbb{G}}}$. Note here that $\left(i\widehat{\beta}_{\tilde{\mathbb{G}}} \, \int  \lambda^{(0)}\cup \d A^{(1)}\right) $ plays a different role than in \eqref{cupwithC}, where the Postnikov was cupped with $C^0$ because $\widehat{\beta}_{\tilde{\mathbb{G}}}$ is strictly speaking valued in $\widehat{\mathcal{A}}_{[-1]}$, and there is no sense in which it can be canceled in the same way a phase can be.  

 We return back to considering the Postnikov under the homomorphism $f$ as in \eqref{FofBeta} and apply this to the partition function of $\cZ_\mathcal{T}$.
By conducting the gauging procedure of \eqref{GaugedPartition} along with the functorial property of $f$ we see that there are mixed anomalies in each subsector labeled by $\widehat{\alpha}$, and the partition function is given by
  \begin{align}\label{2groupgauged}
 \cZ^{\mathcal{T}/\!/\mathcal{A}_{[1]}}_{A_{\widehat{\alpha}}^{(1)};\,C_{\widehat{\alpha}}^{(0)}}=&\int D b_{\widehat{\alpha}}^{(2)}\cZ^{\mathcal{T}}_{A^{(1)}_{\widehat\alpha};\,b^{(2)}_{\widehat\alpha}}\,\exp\left(i\left(\,  C_{\widehat{\alpha}}^{(0)}\cup f(\beta_{\mathbb{G}})_{\widehat{\alpha}} \right)\,\int \lambda^{(0)}_{\widehat\alpha} \cup dA^{(1)}_{\widehat\alpha} \right), \notag\\[2pt]
\end{align}
 where the subscript $\widehat{\alpha}$ denotes the restriction of the gauge field to the subsector $\widehat{\alpha}$ and the full gauge field is written as $C^{(0)} = \bigoplus_{\widehat{\alpha}} C_{\widehat{\alpha}}^{(0)}$.
Here, $\,\langle   C^{(0)}_{\widehat{\alpha}} \cup - \rangle : \hom(\cA_{[1]}\,,\mathbb{C}) \big|_{\widehat{\alpha}}$ and the term $\left(i\,f(\beta_\mathbb{G})_{\widehat{\alpha}} \cup C^{(0)}_{\widehat{\alpha}}\right)$ makes sense as a complex number. This makes\\ $\left(i\,f(\beta_{\mathbb{G}})_{\widehat{\alpha}}\cup C_{\widehat{\alpha}}^{(0)} \right)\int \lambda^{(0)}_{\widehat\alpha} \cup dA^{(1)}_{\widehat\alpha} $ a mixed anomaly.

\subsection{Space of (-1)-form symmetries}
In order to fit in line with the definition of symmetry defects given in \S\ref{startoftwo}, a $(-1)$-form symmetry for a $d$-dimensional theory must be implemented by codimension zero defects, or ``spacefilling defects", which are theories themselves.   This says that in gauging a one-form symmetry, we have effectively projected ourselves onto a subtheory of the original $\mathcal{T}$ existing in the direct sum.  Furthermore, gauging this $(-1)$-form symmetry should give us back a family of theories as can be seen in the following way.  For an element $\widehat{\alpha}$ of $\widehat{\cA}_{[-1]}$ we can build the term $\exp\left( i\int  B^{(2)} \cup C_{\widehat{\alpha}}^{(0)} \right) $ to be inserted into the path integral along with the partition function $\cZ^{\mathcal{T}/\!/ \mathcal{A}_{[1]}}$.  We then take a direct sum over $\widehat{\alpha}$ while integrating over gauge field corresponding to $\widehat{\cA}_{[-1]}$\,,   
\begin{align}
 Z^{\mathcal{T}/\!/ \mathcal{A}_{[1]}/\!/ \widehat{\mathcal{A}}_{[-1]}} &= \bigoplus_{\widehat{\alpha}} \int D c_{\widehat{\alpha}}^{(0)} \int D b^{(2)}\cZ^{\mathcal{T}}_{A^{(1)};\, b^{(2)}}  \exp\left(-i \,\int b^{(2)}\cup c_{\widehat{\alpha}}^{(0)} \right) \notag  \\[2pt]
 &\hspace{75mm}\times \exp\left(i\,\int B^{(2)} \cup c_{\widehat{\alpha}}^{(0)}   \right)\notag\\
 &= \cZ^{\mathcal{T}}_{A^{(1)};\, B^{(2)}}\,,
\end{align}
which takes us back to the original family of theories with a one-form symmetry.

From the point of view of a defect that selects a particular subsector of a theory, we see that there is no such $p$-form symmetry for $p$ $<$ $(-1)$. As a brief note for completeness we give a mathematical way to understand other negative form symmetries.  We explained in \S\ref{startoftwo} that one-form symmetries are one-to-one with group objects in groupoids, such that only $\pi_1$ was nontrivial.  We can define an $n$-groupoid as a category in which objects support $\pi_0, \pi_1,\ldots,\pi_n$ homotopies.  Given $G$, a group object in $n$-groupoids, it is possible to form what we will call $\rB G$, which is an $n+1$-groupoid.  Here, the group law of the $n$-groupoid becomes the composition law in the $n+1$-groupoid.  Furthermore, $\pi_{i-1}G = \pi_i \rB G$, and thus $\pi_{-1}G = \pi_0 \rB G$, the right-hand-side of the equality being well defined.  This gives a view of $(-1)$-form symmetry in terms of homotopy if only $\pi_0 \rB G$ is nontrivial for $G$ is a group object in $(-1)$-groupoids.  Instead of $n$-groupoids being defined with only a single group law, it is possible to include multiple group laws.  In essence, this means that associativity can be given multiple ways of being isomorphic.  Starting with an $n$-groupoid and permitting two group laws, it is possible to form an $(n+1)$-groupoid with a single group law, and subsequently an $(n+2)$-groupoid.  This provides a mathematical way to define $\pi_{-n}$ and therefore ($-n$)-form symmetries if one is willing to consider multiple group laws. 

\subsection{2-group Anomalies}\label{Bosonic2GAnomaly}
In (1+1)$d$ a theory with 2-group global symmetry can itself have an anomaly that is a class in $\rH^3(\mathbb{G}\,,\rU(1))$.  This amounts to asking whether the entire 2-group can be gauged, or if there is an obstruction to doing so.  As was pointed out in \cite{Benini:2018reh}, the fact that this anomaly is a class in the third cohomology is a facet of the dimension of the theory we are considering, and should not be confounded with the Postnikov, which is strictly speaking valued in $\mathcal{A}_{[1]}$ and is not a bona fide anomaly.  We will study this 2-group anomaly by using the Serre spectral sequence and by using the convergence of $\rH^\bullet(G;\rH^\bullet(\cA_{[1]}; \rU(1)))  \Rightarrow \rH^\bullet(\mathbb{G};\rU(1))$; the group of one-form symmetry, in this case, we take to be cyclic of odd order.  The zero-form symmetry we still leave to be a general finite group.  For this and other subsequent calculations, we will only focus on the low degree cohomology. 

The $E_2$ page has $d_2=0$, and the next differential $d_3 = \langle \beta_{\mathbb{G}}, -\rangle$.  If \\
$\omega \in \rH^\bullet(G; \rH^\bullet(\cA_{[1]}; \rU(1)))$, then
\begin{equation}
    d_3(\omega) = \beta_{\mathbb{G}} \cup \omega \in \rH^{\bullet+3}(G; A \otimes \rH^\bullet(\cA_{[1]}\,; \rU(1))).
\end{equation}
To see that this makes sense we note that since $\cA_{[1]}$ is a one-form symmetry, then for the underlying group we have $A = \rH_2(\cA_{[1]}\,;\mathbb{Z}) = \rH_2(K(A,2)\,;\mathbb{Z}) $.  Thus this slant product is the map $\rH_2 \otimes \rH^\bullet \to \rH^{\bullet-2}$. This shows that our claim for $d_3$ is a map $\rH^\bullet(G;\rH^\bullet(\cA_{[1]}; \rU(1))) \to \rH^{\bullet+3}(G;\rH^{\bullet-2}(\cA_{[1]}; \rU(1)))$, as it should be.
The $E_2$ page in degree $\leq 4 $ looks like 
 \begin{equation}\label{E2pageof2Group}
    \begin{array}{c|cccccc}
     
     S^2 \widehat{A}&S^2 \widehat A & \rH^1(G; S^2\widehat A) & \rH^2(G; S^2\widehat A) & \rH^3(G; S^2\widehat{A})& \rH^4(G; S^2\widehat A)\\
     0&0 & 0 & 0 & 0 & 0 \\
     \widehat A&\widehat A &  \rH^1(G;\widehat A) & \rH^2(G;\widehat A) & \rH^3(G;\widehat A)& \rH^4(G;\widehat A)\\
     0&0 & 0 & 0 & 0 & 0 \\
     \rU(1)&\rU(1) & \rH^1(G;\rU(1)) & \rH^2(G;\rU(1)) & \rH^3(G;\rU(1))& \rH^4(G;\rU(1))\\ \hline
       &0 & 1 & 2 & 3 & 4 \,\,\,\,\quad  ,
    \end{array}
\end{equation}
where $\widehat{A} = \rH^2(\mathcal{A}_{[1]}; \rU(1))$, and $S^2 \widehat{A} = \Sym^2 \widehat{A}$ which are the quadratic forms on $\widehat{A}$.
In total degree 3, we have $\bigoplus_{i=0}^3 \rH^i(G\,; \rH^{3-i}(\cA_{[1]} \, , \rU(1))$ and so we consider if any of these elements can support, or receive a differential $d_3$.  By the property of differential 
\begin{equation}
    \rH^i(G; \rH^{2-i}(\cA_{[1]}\,;\rU(1))) \overset{d_3}\to \rH^{i+3}(G; \rH^{-i}(\cA_{[1]}\,;\rU(1))) = 0 , \hspace{2mm}\text{if } i \neq 0,
\end{equation}
and therefore the only $d_3$ in this case is from $\rH^0(G; \rH^2(\cA_{[1]}\,;\rU(1))) = \widehat A$, which lands in $\rH^3(G; \rH^0(\cA_{[1]}\,;\rU(1))) = \rH^3(G; \rU(1))$.  Furthermore we have 
\begin{equation}
    \rH^i(G; \rH^{3-i}(\cA_{[1]}\,;\rU(1))) \overset{d_3}\to \rH^{i+3}(G; \rH^{1-i}(\cA_{[1]}\,;\rU(1))) = 0 , \hspace{2mm}\text{if } i \neq 0,1\,,
\end{equation}
but the coefficient $\rH^3(\cA_{[1]}\,;\rU(1)) = 0$.  Therefore the only $d_3$ here is from \\
$\rH^1(G\,,\rH^2(\cA_{[1]}\,;\rU(1))$ to $\rH^4(G\,;\rU(1))$.  The $E_\infty$ page in total degree 3 contains two entries, which are
\begin{equation}
    \ker(d_3 = \langle -\cup\beta\rangle : \H^1(G;\widehat A) \to \H^4(G;\rU(1))) \text{ and } \operatorname{coker}(\d_3 : \widehat A \to \H^3(G;\rU(1)))\,.
\end{equation}
There is an extension problem to solve here, with $\rH^3(\mathbb{G}\,;\rU(1))$ fitting in the short exact sequence 
\begin{equation}
    \operatorname{coker}(\d_3 : \widehat A \to \H^3(G;\rU(1))) \to \rH^3(\mathbb{G}\,,\rU(1)) \to \ker(d_3 = \langle -\cup\beta_{\mathbb{G}}\rangle : \H^1(G;\widehat A) \to \H^4(G;\rU(1))).
\end{equation}
There is an image of $\omega \in \rH^3(\mathbb{G}; \rU(1))$, which we call $\alpha$, is such that $d\alpha = 0$ and is in the kernel of $d_3$, i.e., is zero in cohomology.  Since $\beta_{\mathbb{G}}$ was chosen to be a cocycle, then $\alpha \cup \beta_{\mathbb{G}}$ is a cocycle; we claim that it is $\d \gamma$ for some 3-cochain $\gamma$ coming from the cokernal set, that witnesses $\alpha$ being in the kernal of $d_3$.  This means that the cohomology  $\rH^3(\mathbb{G};\rU(1))$ consists of pairs $(\alpha,\gamma)$ where $\alpha : G \to \widehat A$ is a homomorphism, and $\gamma \in \rC^3(G;\rU(1))$, such that 
\begin{equation}
     \d \gamma = \alpha \cup \beta_{\mathbb{G}} .
\end{equation}
 If we go to a specific case where $G = \bZ_2$, then $\alpha = 0$ since $A$ was a finite group of odd order, and then $\gamma$ is in fact a cocycle.  In this case the group the cohomology would just be given by $\gamma \in  \rH^3(G,\rU(1))= G = \bZ_2$.  This is also true in general whenever $G$ is finite cyclic, and  gcd($|G|$,\,$|A|$) = 1.
\subsection{2-groups and Supercohomology}\label{fermioic2GAnomaly}
We calculate the anomalies in (1+1)$d$ fermionic theories with 2-group global symmetry; these anomalies live in supercohomology. We will more specifically consider what is known as extended supercohomology \cite{Wang:2017moj} in the following. Bosonic anomalies that are $\tfrac{1}{2}\,$(mod 1), when restricted to a one-form symmetry subgroup of $\mathbb{G}$, become trivialized in supercohomology.

We take fGP$^\times$ to be the spectrum of fermionic phases.  This is a sequence of of topological spaces, namely fermionic invertible gapped systems $\ldots, \fGP^{\times}_{-1},\, \fGP^{\times}_{0},\, \fGP^{\times}_{1},\\
\,\fGP^{\times}_{2},\ldots$ where the subscript denotes the spacetime dimension.  This sequence comes with homotopy  equivalences $\fGP^{\times}_{n-1} \overset{\sim}{\to} \Omega \fGP^{\times}_n $, and since $\pi_n \Omega \fGP^{\times}_n = \pi_{n+1} \fGP^{\times}_n$, we define the homotopy groups of $\fGP^{\times}$ by $\pi_n \fGP^{\times} = \pi_0 \fGP^{\times}_{-n}$ and $\pi_{-n} \fGP^{\times} = \pi_{0}\fGP^{\times}_{n} $.   Calculating these groups gives the $n$-dimensional fermionic phases with abelian group structure $\pi_0 \fGP^{\times}_{-n}$, where the group composition is by stacking. In what follows we will only focus on the low dimensional homotopy groups of this spectrum, the well established ones are \cite{Gaiotto_2019}:
\begin{equation}
     \pi_0 \fGP^\times = \rU(1), \quad \pi_{-1} \fGP^\times = \bZ_2,\quad \pi_{-2} \fGP^\times = \bZ_2.
\end{equation}
The only nontrivial degree-2 stable cohomology operation from $\bZ_2$ to $\rU(1)$ is $(-1)^{\Sq^2}$, as $\Sq^2: \rH^\bullet(-\,; \bZ_2) \to \rH^{\bullet+2}(-\,; \bZ_2) $ and $(-1)^x: \rH^\bullet(- \,; \bZ_2 ) \to \rH^\bullet(-\,;\rU(1))$. The nontrivial degree-2 stable cohomology operation connecting $\bZ_2$
to $\bZ_2$ is just $\Sq^2$. 

An $n$-cocycle in supercohomology ($\SH^n$) consists of a triple $(\alpha,\beta,\gamma)$  where $\alpha$ is a degree-$n$ $\rU(1)$-cochain, $\beta$ is a degree-$(n-1)$ $\bZ_2$-cochain, and $\gamma$ is a degree-$(n-2)$ cochain and they solve:
\begin{equation}
\label{rSH-cocycle}
     \d\gamma = 0, \qquad \d\beta = \Sq^2\gamma, \qquad \d\alpha = (-1)^{\Sq^2 \beta} + f(\gamma).
\end{equation}
In the literature, $\gamma$ is referred to as the ``Majorana layer" when in degree one, and $\beta$ is the ``Gu-Wen" layer when in degree two, and $\alpha$ is a 't Hooft anomaly in the bosonic sense.  We want to calculate the supercohomology $\SH^3(\mathbb{G})$ for $\mathbb{G} = \mathcal{A}_{[1]}\rtimes_{\beta_{\mathbb{G}}} G$, but now $G$ and $\cA_{[1]}$ are the group $\mathbb{Z}_2$ and $\rB\bZ_2$ respectively, for otherwise supercohomology reduces to standard cohomology.

For completeness we present this calculation in pieces, where we also calculate the supercohomology of $\mathbb{G} = \bZ_2$ and $\mathbb{G} = \rB\bZ_2$.  The supercohomology as a generalized cohomology theory takes value in a spectrum.  By using the Atiyah-Hirzebruch spectral sequence we have $\SH^\bullet(\bZ_2) \Leftarrow \rH^\bullet(\bZ_2, \SH^\bullet(\pt))$ where $\SH^\bullet(\pt) = \pi_{-\bullet}(\pt)$.  In the extended cohomology case we need:
\begin{equation}
    \SH^0(\pt) = \rU(1)\,, \quad \SH^1(\pt) = \bZ_2\,, \quad \SH^2(\pt) = \bZ_2\,,
\end{equation}
and the homotopy in higher nonnegative cohomological degree vanishes.  The ring $\rH^\bullet(\bZ_2;\bZ_2) = \bZ_2[t]$ with $t$ in degree one, and except in degree $0$, the map $(-1)^{x} : \bZ_2 \to \rU(1)$ is a surjection on $\rH^\bullet(\bZ_2;\bZ_2) \to \rH^\bullet(\bZ_2; \rU(1))$.
This gives the $E_2$ in low degree as
\begin{equation}
    \begin{array}{c|ccccc}
    \bZ_2 & \bZ_2 & \bZ_2t & \bZ_2t^2 & \bZ_2t^3 & \bZ_2 t^4\\
    \bZ_2 & \bZ_2 & \bZ_2t & \bZ_2t^2 & \bZ_2 t^3 & \bZ_2t^4 \\
    \rU(1) &\rU(1) & (-1)^{\bZ_2t} & 0 & (-1)^{\bZ_2t^3} & 0 \\ \hline
    & 0 & 1 & 2 & 3 & 4 \,\quad\,.
    \end{array}
\end{equation}
In what follows we will strip off the $\bZ_2$ for simplicity and only leave the generator. The $d_2$ differential in the Atiyah-Hirzebruch spectral sequence is the $k$-invariants of the spectrum, these are $\Sq^2 : E_2^{\bullet,2} \to E_2^{\bullet+2,1}$ and $(-1)^{\Sq^2} : E_2^{\bullet,1} \to E_2^{\bullet+2,0}$.  On generators $t$, $\Sq^2$ acts as a second order operator by differentiation.  Namely, $\Sq^2 = t^4 \frac{1}{2} \frac{\p^2}{\p ^2t}: t^i \mapsto \binom{i}{2} t^{i+2}$.   The $d_2$ mapping out of generators in $i \equiv 0,1 \mod 4$ therefore vanish, and the $d_2$ mapping out of generators in $i \equiv 2,3 \mod 4$ are isomorphisms.  The $E_3$ page is \begin{gather}
    \begin{array}{c|ccccccccc}
    \bZ_2 & 1 & \quad t & \quad 0 & \quad0 &  \quad t^4 & \\
    \bZ_2 & 1 & \quad t & \quad t^2 &  \quad0 & \quad0 &  \\
    \rU(1) &\rU(1) & \quad t & \quad 0 & \quad t^3 & \quad0 \\ \hline
    & 0 & 1 & 2 & 3 & 4\,\quad . 
    \end{array}
\end{gather}
In total degree $\leq 4$, the above $E_3$ page is the $E_\infty$ page, and so in total degree three  $\SH^3(\bZ_2) = \bZ_2.\bZ_2.\bZ_2$; now we are left with an extension problem to solve. 

We first look at the extension $\bZ_2.\bZ_2$ between the top and middle row.  The extension gives information about the failure of $\Sq^2$ to act linearly on cocycles.  For $a,b \in E^{1,2}_\infty$, $\Sq^2(a+b) =(a+b)\cup(a+b)= a^2+b^2$ only if $a\,\cup\, b = b\, \cup \,a$, i.e., that the cup product is commutative.  The lack of commutativity gives an element $a\cup b \in E^{2,1}_\infty$, so the extension is nontrivial and we get a $\bZ_4$ from the top and middle rows.   To understand the extension of $\bZ_4.\bZ_2$ we consider the image of $(-1)^{{\Sq^2} (t^2)}$ which as shown on the $E_3$ page is zero in cohomology, implying that it is a coboundary $d\lambda$ for $\lambda \in \rH^3(\bZ_2;\rU(1)) = E^{3,0}_{\infty}$.  The extension information is therefore embedded in $\lambda$, much like how the data for the first extension was embedded in $a\cup b$, and so the extension is nontrivial.  We see that  $\SH^3(\bZ_2) = \bZ_8$.  It is also possible to arrive at this conclusion on a more physical level;  the spectral sequence reveals that in total degree three, whatever group this is, must have order eight.  The supercohomology $\SH^\bullet(K(\bZ_2,1))$ classifies the superfusion categories with $\bZ_2$ fusion rules. The bosonic shadow must then have an object akin to a ``fermion" with $\bZ_2$ fusion rules, and there are eight categories:
four with Ising fusion rules, two with $\bZ_4$ fusion rules, one with $\bZ_2^2$ fusion rules and nontrivial associator, and one with $\bZ_2^2$ fusion rules and trivial associator,
which are part of the $\Spin(N)_1$ monoidal categories.   Recognizing them as such allows us to recognize the group structure as a $\bZ_8$, and furthermore allows for the identification of the $\bZ_8$ in supercohomology with the $\bZ_8$ of Bott periodicity. 

We now compute $\SH^\bullet(\rB\bZ_2)$ by converging from $\rH^\bullet(\rB\bZ_2\,;\SH^\bullet(\pt))$.  The cohomology ring of $\rB\bZ_2$ with coefficients in $\bZ_2$ is generated over the Steenrod algebra with a generator $T$ in degree two \cite{hatcher2002algebraic}:
\begin{equation}
    \rH^\bullet(\rB\bZ_2\,,\bZ_2) = \bZ_2[T\,,\,\Sq^1( T)\,,\, \Sq^2 (T)\,,\, \Sq^2\Sq^1(T)\,,\, \Sq^4\Sq^2\Sq^1 (T)\,,\ldots]\,.
\end{equation}
The $E_2$ page is 
  \begin{equation}\label{superofBZ2}
    \begin{array}{c|ccccc}
    \bZ_2 & 1 & 0 & T & \Sq^1(T) & \Sq^2 (T)\\
    \bZ_2 & 1 & 0 & T & \Sq^1(T) & \Sq^2 (T)\\
    \rU(1) &\rU(1)&0 & \bZ_2 & 0 & \bZ_4  \\ \hline
    & 0 & 1 & 2 & 3 & 4 \,\quad.
    \end{array}
\end{equation}
We want to determine if the $d_2$ mapping out from $T$ is nonzero.  The Universal Coefficient theorem gives the cohomology $\rH^\bullet(A, \rU(1))$ as an extension of a hom class and an Ext class in homology with $\bZ$ coefficients.  This reveals that row zero of the $E_2$ page consists solely of hom classes, because $\Ext(A,\rU(1)) = 0$ for any abelian group $A$.  The first row consists of hom and Ext classes; applying $\Sq^2$ to $T$ in $E^{2,1}_2$ gives $\Sq^2(T)$ but there is no Ext class in that degree, which mean $\Sq^2(T)$ must be a hom class.  Since the map $(-1)^x : \bZ_2 \to \rU(1)$ is injective, the map on cohomology must be injective on hom classes.  Therefore $(-1)^{\Sq^2(T)} \neq 0$, and because this map is injective, we find that $T$ is killed by this differential.  Thus, $\SH^3(\rB\bZ_2) = 0$ as nothing survives in that degree.

At this point we can compute $\SH^\bullet(\rB\bZ_2 \rtimes_\beta \bZ_2)$, on the $E_2$ page for the Atiyah-Hirzebruch spectral sequence we will need $\rH^\bullet(\rB\bZ_2 \rtimes_\beta \bZ_2\,;\rU(1))$ and $\rH^\bullet(\rB\bZ_2 \rtimes_\beta \bZ_2\,;\bZ_2)$.  We build up the $E_2$ page for $\SH^\bullet(\rB\bZ_2 \rtimes_\beta \bZ_2)$ in steps, first starting by obtaining $\rH^\bullet(\rB\bZ_2 \rtimes_\beta \bZ_2\,;\rU(1))$ and $\rH^\bullet(\rB\bZ_2 \rtimes_\beta \bZ_2\,;\bZ_2)$ with the Serre spectral sequence. 
The $E_2$ page of $\rH^\bullet(\bZ_2\,;\rH^\bullet(\rB\bZ_2\,;\rU(1)))$ is 
\begin{equation}
    \begin{array}{c|ccccc}
    \bZ_4 & \bZ_4 & \bZ_2 & \bZ_2 & \bZ_2 & \bZ_2\\
    0&0&0&0&0&0\\
    \bZ_2 & \bZ_2 & \bZ_2 & \bZ_2 & \bZ_2 & \bZ_2\\
    0 & 0 & 0 & 0 & 0 & 0\\
    \rU(1) &\rU(1)&\bZ_2 & 0 & \bZ_2 & 0  \\ \hline
    & 0 & 1 & 2 & 3 & 4 \,\quad.
    \end{array}
\end{equation}
The $d_2$ differential vanishes for degree reasons and the $d_3$ differential is given by information of the extension, $\beta \in \rH^3(\bZ_2\,;\rB\bZ_2)$.  The row in degree two represents $\widehat{\bZ}_2 = \hom(\rB\bZ_2,\rU(1))$ by the Hurewicz theorem, which gives $d_3 =  (-1)^{\langle - \cup \beta \rangle} : \rH^\bullet(\bZ_2\,;\bZ_2) \to \rH^\bullet(\bZ_2\,;\rU(1))$.  We now consider the $d_3$ map in the row of degree four, where the $\bZ_4$ denotes the space of quadratic forms on $\bZ_2$.  We must therefore have a map $\{\text{quadratic forms}\}\otimes \bZ_2 \to \widehat{\bZ}_2$, which in this case is given by the mod 2 reduction, and we find 
\begin{align}
    d_3: \,\rH^\bullet(\bZ_2 \,; \bZ_4) &\to \rH^\bullet(\bZ_2\, ;\bZ_2)\notag\\[2pt] 
        & x \mapsto xt^3 \mod2.
\end{align}
The $E_4$ page converges to the $E_\infty$ page in low degree and is 
\begin{equation}
    \begin{array}{c|ccccc}
    \bZ_4 & \bZ_2 & \bZ_2 & 0 & \bZ_2 & 0\\
    0&0&0&0&0&0\\
    \bZ_2 & 0 & \bZ_2 & 0 & 0 & 0\\
    0 & 0 & 0 & 0 & 0 & 0\\
    \rU(1) &\rU(1)&\bZ_2 & 0 & 0 & 0  \\ \hline
    & 0 & 1 & 2 & 3 & 4 \,\quad
    \end{array}
\end{equation}
with $\rH^\bullet(\rB\bZ_2 \rtimes_\beta \bZ_2\,;\rU(1)) = \rU(1)\,,\bZ_2\,,0\,,\bZ_2\,,\bZ_2\,,\ldots$

For the calculation of $\rH^\bullet(\rB\bZ_2 \rtimes_\beta \bZ_2\,;\bZ_2)$, on the $E_2$ page we drop the $\bZ_2$ everywhere, and only present the generators at that total degree.
This gives
\begin{equation}
    \begin{array}{c|ccccc}
    \Sq^2 T & \quad \Sq^2 T & \quad t\Sq^2 T & \quad t^2\Sq^2 T & \quad t^3\Sq^2 T & \quad t^4\Sq^2 T\\
    \Sq^1 T& \quad \Sq^1 T & \quad t\Sq^1 T &\quad t^2 \Sq^1 T&\quad t^3\Sq^1T &\quad t^4 \Sq^1T\\
    T & T & tT &\quad t^2T &\quad t^3T &\quad t^4T\\
    0 & 0 & 0 & 0 & 0 & 0\\
    1 &1&t & t^2& t^3 & t^4  \\ \hline
    & 0 & 1 & 2 & 3 & 4 \,\quad,
    \end{array}
\end{equation}
with $d_3 = t^3 \frac{d}{d T}$.   In low degree we have $\rH^\bullet(\rB\bZ_2 \rtimes_\beta \bZ_2\,;\bZ_2) =  \bZ_2, \bZ_2, \bZ_2, \bZ_2, \bZ_2^2, \dots$.   This is consistent with $\rH^\bullet(\rB\bZ_2 \rtimes_\beta \bZ_2\,;\rU(1)) = \rU(1)\,,\bZ_2\,,0\,,\bZ_2\,,\bZ_2\,,\ldots$, where by the Universal Coefficient theorem, each $\bZ_2$ is a class of the same degree, and a class in one degree lower.  On $E_\infty$, a basis for $\rH^\bullet(\rB\bZ_2 \rtimes_\beta \bZ_2\,;\bZ_2)$ is $\{1, t, t^2, \Sq^1 T, T^2, t \Sq^1 T, \dots\}$, and the ring structure is $ \bZ_2[t,\Sq^1T, T^2, \dots]/ (t^3=0,\dots)$.  We assemble now the $E_2$ page for supercohomology  
\begin{equation}
    \begin{array}{c|ccccc}
    \bZ_2 & \quad \bZ_2 1 &\quad \bZ_2 t & \quad \bZ_2 t^2 &\quad \bZ_2 \Sq^1 T & \quad \bZ_2.\bZ_2 \,T^2\\
    \bZ_2 & \quad \bZ_2 1 & \quad \bZ_2 t &\quad \bZ_2 t^2 &\quad \bZ_2 \Sq^1 T & \quad \bZ_2.\bZ_2 \,T^2\\
    \rU(1) &\quad \rU(1)&\quad \bZ_2 & 0& \bZ_2 & \bZ_2  \\ \hline
    & 0 & 1 & 2 & 3 & 4 \,\quad.
    \end{array}
\end{equation}
We notice that the bottom row on $E_\infty$ is the image of $\rH^\bullet( \bZ_2\,;\rU(1)) \to \rH^\bullet(\rB\bZ_2 \rtimes_\beta \bZ_2\,;\rU(1) )$, therefore nothing on that row survives beyond degree higher than three, and the $d_2$ differential, $\Sq^2$ kills the $t^2$ in degree $E^{2,1}_2$ as $\Sq^2(t^2) = t^4=0$.   Finally, $t$ in $E^{1,2}_2$  survives because $\Sq^2$ vanishes there.   Altogether, this suggest that the anomaly for a fermionic theory in (1+1)-dimensions with 2-group global symmetry is $\bZ_4$. \footnote{It is also possible to calculate this supercohomology by converging to it with $\rH^\bullet(G\,,\SH^\bullet(\rB\bZ_2))$.}
\section{Split 2-groups and Symmetry Fractionalization}\label{symfracsection}
\subsection{Review of Symmetry Fractionalization}\label{revsymfrac}
We now consider the case in which the functorial obstruction class $\beta_{\mathbb{G}}$ to the extension of $G$ by $\mathcal{A}_{[1]}$ is trivial in $\rH^3(G;\mathcal{A}_{[1]})$.   Since this is a special case of a 2-group we will call this case a ``split 2-group".  With an action $\rho: G \to \operatorname{Aut}(\mathcal{A}_{[1]})$ and trivial Postnikov, then the extension $\mathbb{E}$  of $G$ by $\mathcal{A}_{[1]}$ inducing $\rho$ is in bijection with classes $\nu(\textbf{g},\textbf{h}) \in \rH^2_{\rho}(G;\mathcal{A}_{[1]})$.  Our split 2-group is known in the literature as a symmetry fractionalized phase, which is specified by $\nu$ \cite{Hsin:2019gvb,Barkeshli:2014cna,Chen_2017}.  For (2+1)$d$ phases, one can
understand symmetry fractionalization as the difference between an anyon $a$ being acted upon by $\textbf{g}$ and $\textbf{h}$ defects separately, versus being acted upon by the composite $\textbf{gh}$ defect. The relationship between different symmetry fractionalizations also becomes clear from a physical point of view when we consider modifying the junction between three zero-form defects to include an anyon $\alpha$. 

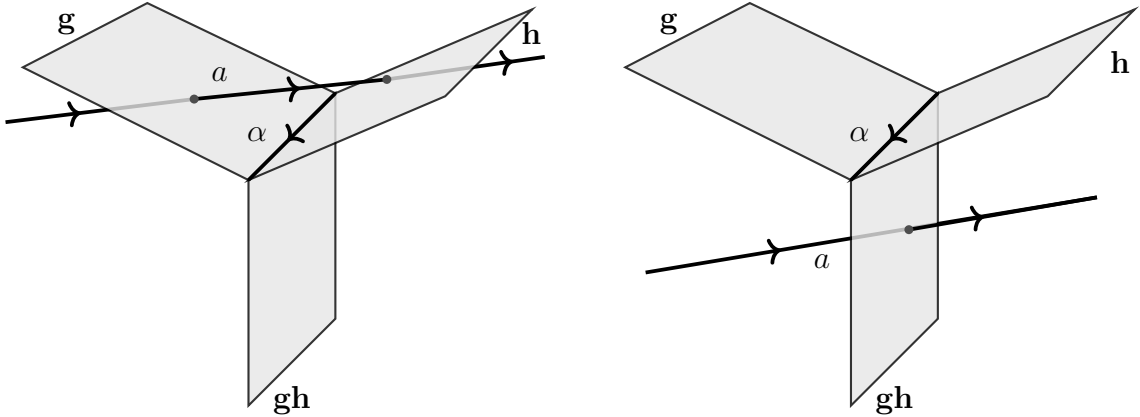
\begin{figure}[ht]
    \begin{center}
    \begin{tikzpicture}[thick]
      \def\Depth{4}
        \def\Height{2}
        \def\Width{2}
        \coordinate (O) at (0,0,0);
        \coordinate (A) at (0,\Width+1,0);
        \coordinate (B) at (0,\Width+1,\Height+1);
        \coordinate (C) at (0,0,\Height+1);
        \coordinate (D) at (\Depth,0,0);
        \coordinate (E) at (\Depth,\Width,0);
        \coordinate (F) at (\Depth,\Width,\Height);
        \coordinate (G) at (\Depth,0,\Height);
            \draw[decoration={markings, mark=at position 0.4 with {\arrow{>}}}, postaction={decorate},line width= .5mm](-4,\Width+1,\Height/2) -- (-1.3,\Width+1.5,\Height/2+.5); 
         
     \draw[black,fill=black!10,opacity=0.8] (O) -- (A) -- (B) -- (C) -- cycle;
     
    \draw[decoration={markings, mark=at position 0.8 with {\arrow{>}}}, postaction={decorate},line width= .5mm]
    (1.3,\Width+1.8,\Height/2+.6) -- (3.4,\Width+2.1,\Height/2+.6); 

     \draw[black,fill=black!10,opacity=0.8] (A) -- (B) -- (3,4.5,4) -- (3,4.5,1)--cycle;

     \draw[black,fill=black!10,opacity=0.8] (A) -- (B) -- (-3,4.5,3) -- (-2.5,4.2,0) -- cycle;
     
      \draw[decoration={markings, mark=at position 0.55 with {\arrow{>}}}, postaction={decorate},line width= .5mm] (-1.3,\Width+1.5,\Height/2+.5)-- (1.3,\Width+1.8,\Height/2+.6);
      
       \tikzstyle{vertex}=[circle,fill=black!70, minimum size=.05pt,inner sep=1.2pt]
    \node[vertex] at (-1.3,\Width+1.5,\Height/2+.5) {};
    
    \tikzstyle{vertex}=[circle,fill=black!70, minimum size=.05pt,inner sep=1.2pt]
    \node[vertex] at (1.3,\Width+1.8,\Height/2+.6) {};

     \draw[decoration={markings, mark=at position 0.55 with {\arrow{>}}}, postaction={decorate},line width =.5mm] (A) -- (B);
     
      \draw[black] (-3,4.5,1.5) node {$\textbf{g}$};
    \draw[black] (2.8,4,.5) node {$\textbf{h}$};
    \draw[black] (0,-.5,1.5) node {$\textbf{gh}$};
     \draw[black] (0.5,4,4) node {$\alpha$};
     \draw[black] (0,4.8,4) node {$a$};
       \end{tikzpicture}
      \label{lineup}
\qquad 
    \begin{tikzpicture}[thick]
      \def\Depth{4}
        \def\Height{2}
        \def\Width{2}
        \coordinate (O) at (0,0,0);
        \coordinate (A) at (0,\Width+1,0);
        \coordinate (B) at (0,\Width+1,\Height+1);
        \coordinate (C) at (0,0,\Height+1);
        \coordinate (D) at (\Depth,0,0);
        \coordinate (E) at (\Depth,\Width,0);
        \coordinate (F) at (\Depth,\Width,\Height);
        \coordinate (G) at (\Depth,0,\Height);
            \draw[decoration={markings, mark=at position 0.3 with {\arrow{>}}}, postaction={decorate},line width= .5mm](-3.5,\Width-1,\Height/2) -- (2.5,\Width,\Height/2);
     \draw[black,fill=black!10,opacity=0.8] (O) -- (A) -- (B) -- (C) -- cycle;
     \draw[black,fill=black!10,opacity=0.8] (A) -- (B) -- (3,4.5,4) -- (3,4.5,1)--cycle;
     \draw[black,fill=black!10,opacity=0.8] (A) -- (B) -- (-3,4.5,3) -- (-2.5,4.2,0) -- cycle;
     \draw[decoration={markings, mark=at position 0.55 with {\arrow{>}}}, postaction={decorate},line width = .5mm] (A) -- (B);
     \draw[decoration={markings, mark=at position 0.4 with {\arrow{>}}}, postaction={decorate},line width= .5mm]  (0,1.57,\Height/2)  -- (2.5,\Width,\Height/2);
      \tikzstyle{vertex}=[circle,fill=black!70, minimum size=.05pt,inner sep=1.2pt]
    \node[vertex] at (0,1.57,\Height/2) {};
    
    \draw[black] (-3,4.5,1.5) node {$\textbf{g}$};
    \draw[black] (3,4,1.5) node {$\textbf{h}$};
    \draw[black] (0,-.5,1.5) node {$\textbf{gh}$};
     \draw[black] (0.5,4,4) node {$\alpha$};
      \draw[black] (0,2.3,4) node {$a$};
   \end{tikzpicture}
   \caption{The line $a$ is acted on by the zero-form symmetry as it punctures the surfaces, and passing the line $a$ by $\alpha$ results in a braiding.}
   \label{linedown}
   \end{center}
\end{figure}

\label{ga}

A line operator implementing the one-form symmetry when passing from above $\alpha$ to below in figure \ref{linedown}, picks up a braiding \cite{Barkeshli:2019vtb}.  If we further choose the class of $[0] \in \rH^2(G;\mathcal{A}_{[1]})$ then this means the symmetries of the phase is simply $ \cA_{[1]}\times G$, i.e, there is no action of the zero-form  symmetry on the one-form symmetry.  The above information can also be presented by considering the following exact sequence 
 \begin{equation*}
     \begin{tikzcd}
    0\arrow{r} & \mathcal{A}_{[1]}\arrow{r} & \mathbb{E}\arrow{r} & G\arrow{r}\arrow[bend left=33]{l}{\varphi} & 0 \, , \\
\end{tikzcd}
 \end{equation*}
where the splitting map $\varphi$ gives the trivial class $\beta_{\mathbb{G}}$, with $\mathbb{E} = \mathcal{A}_{[1]}\rtimes G$.   There are still $\rH^2(G \,; \mathcal{A}_{[1]})$ choices of conjugacy classes in how we trivialize, for if $\varphi_1$ and $\varphi_2$ are two such splittings, then there could exist $m\in \mathcal{A}_{[1]}$ such that $\varphi_2(g) = (1,m)\,\circ \varphi_1(g)\, \circ (1,m)^{-1}$ in $\mathbb{E}$.   In (1+1)$d$ it makes sense to consider $\nu \in \rH^2(G\,;\mathbb{C}[\mathcal{A}_{[1]}]^\times)$, just as we did for the case of the 2-group.  By the same argument as we gave for the Postinikov, we can write $f(\nu) \in \bigoplus_{\hat{A}}\, \rH^2(G\,;\rU(1))$.  The symmetry fractionalization thus appears as if we are choosing to assign a  (1+1)$d$ zero-form discrete torsion \cite{Vafa:1994rv} for the different disjoint theories.

\subsection{Anomalies for the Split 2-group}

Anomalies in symmetry fractionalized theories in (1+1)$d$ are classified by a class in $\rH^3(\mathcal{A}_{[1]}\rtimes G\,;\rU(1))$.  We first remark that if $\nu = [0]$, then $\rho: G\to \operatorname{Aut}(\mathcal{A}_{[1]})$ is trivial and the cohomology $\rH^3_\rho(\cA_{[1]}\times G\,;\rU(1))$ is calculated with the Kunneth formula.   That is to say that all higher differentials in the spectral sequence vanish and $E_2 = E_\infty$. If $\nu$ is not trivial, then the $d_2$ differential is given by cupping with $\nu$, i.e. $\langle \nu  \cup \, -\rangle$. The $E_2$ page of the spectral sequence we run in order to calculate the anomalies takes the form of \eqref{E2pageof2Group}, and has no  elements that can receive a $d_2$ differential.  The cohomology $\rH^3(\mathcal{A}_{[1]}\rtimes G\,;\rU(1))$ fits in a short exact sequence
\begin{equation} \label{symfracSES}
    \rH^3(G\,;\rU(1))  \to \rH^3(\mathcal{A}_{[1]}\rtimes G\,;\rU(1))\to  \rH^1(G\,;\widehat A).
\end{equation}
The first map is given by the pullback along the map $\mathbb{E}\to G$ (given by projection).  The second map can be built in the following way: we begin by considering  a function from $G$ to 2-cochains, $\delta$, on $\mathcal{A}_{[1]}$ that goes 
\begin{equation}
\Delta(g) = \delta g - \delta,\, \,\text{for}\,\, d\delta \in \rH^3(\mathcal{A}_{[1]}\,;\rU(1))\,. 
\end{equation}
Next, we  prove properties of $\Delta$.  We note that $d \delta$ can be thought of as the restriction of a class $\omega \in \rH^3(\mathcal{A}_{[1]}\rtimes G\,;\rU(1))$ to $\rH^3(\mathcal{A}_{[1]}\,;\rU(1))$ because away from the prime 2, $\rH^3(\mathcal{A}_{[1]}\,;\rU(1))=0$.
By taking the  differential in $\mathcal{A}_{[1]}$ of $\Delta$ we see that 
\begin{equation}
    d_{\mathcal{A}_{[1]}} \Delta (g) = d \delta\, g - d\delta = 0\,,
\end{equation}
and by taking the twisted differential in $G$ we get 
\begin{align} 
    d_{G} \Delta(g,h) &=  \Delta(g) h  - \Delta(gh)+\Delta(h) \notag \\
     & = (\delta g - \delta )h - (\delta gh - \delta) + \delta h - \delta \notag \\
     &=0\,.
\end{align}
This shows that $\Delta(g)$ is actually a 2 cocycle in $\mathcal{A}_{[1]}$ and a one cocycle in $G$, i.e $\Delta \in \rH^1(G \,; \rH^2(\cA_{[1]}; \rU(1))) = \rH^{1}(G\,;\widehat{\mathcal{A}}_{[-1]})$.  A different choice of $\delta$, differing by another cochain $\chi \in \rH^2(\cA_{[1]}\,; \rU(1) )$, changes $\Delta(g)$ to $\Delta(g) + \chi \,g - \chi$.  The difference $\chi \,g - \chi = d_G \chi$, is the differential of a zero-cochain valued in $\rH^1(G; \widehat{\cA}_{[1]})$.  We see that the short exact sequence in \eqref{symfracSES} does not depend on splitting of $\mathbb{E}\to G$ and therefore for each splitting, there is an isomorphism
\begin{equation}\label{SymFracAnomaly}
    \rH^3(\mathcal{A}_{[1]}\rtimes G\,;\rU(1)) \cong \rH^3(G\,;\rU(1)) \oplus \rH^1(G\,;\widehat{\cA}_{[-1]})\, ,
\end{equation}
where the direct sum on the right hand side can be evaluated simply in our case, yielding the anomaly. 
   Since two splittings differ by $\nu$, a change in a splitting by $\nu$ changes the direct sum side of the isomorphism by 
\begin{equation}
    \begin{pmatrix} 1 &\quad  \langle \nu \cup - \rangle  \\ 0 & 1 \end{pmatrix},
\end{equation}
where explicitly $\langle \nu \cup - \rangle$ is a map from $\rH^1(G;\widehat{\cA}_{[-1]}) \to \rH^3(G\,;\rU(1))$.
 
\subsection{Supercohomology for the Split 2-group}
In the case of calculating supercohomology for the group $\mathbb{E} =\mathcal{A}_{[1]}\rtimes G $, this semi-direct product reduces to a product because there are no automorphism of $\bZ_2$.  As in the previous section, away from the prime $2$ this is just regular cohomology.  The spectral sequence in this case gives
$\SH^\bullet(\mathcal{A}_{[1]}\times G) \Leftarrow  \rH^\bullet( G \,; \SH^\bullet(\mathcal{A}_{[1]}))\cong \H^\bullet(G\,;\bZ)\otimes^{\mathbb L} \SH^\bullet(\mathcal{A}_{[1]})$ \footnote{The reader is reminded that the notation $$\bigoplus_{i+j=n} \rH^i(G\,;M) \otimes^{\mathbb L} \rH^j(G'\,;M) = \bigoplus_{i+j=n} \rH^i(G\,;M) \otimes \rH^j(G'\,;M)  \bigoplus_{i+j=n+1} \Tor(\rH^i(G\,;M)\,,\rH^j(G'\,;M))$$ }, where $\mathbb L$ denotes the left derived tensor product.  The above isomorphism is given by the Universal Coefficient theorem.  This implies $\SH^3(\rB\bZ_2\times \bZ_2) \Leftarrow \bigoplus_{i+j=3}  \H^i(\bZ_2\,;\bZ)\otimes^{\mathbb L} \SH^j(\rB\bZ_2) $.
%
%
In order to evaluate the Tor group up to this degree, we in principle need $\SH^4(\rB\bZ_2)$.  One could bypass this calculation due to the fact that $\rH^0(\bZ_2\,;\bZ) = \bZ$ and $\Tor(A\,,\bZ)=0$ for any finite abelian group $A$.  We nonetheless present this calculation because this allows us to say some facts about braided fusion supercategories with $\bZ_2$ fusion rules, since these are parametrized by $\SH^4(\rB\bZ_2)$.  We have the $E_2$ page of $\SH^\bullet(\rB\bZ_2)$ in low degrees as 
\begin{equation}
    \begin{array}{c|ccccc}
    \bZ_2 & 1 & 0 & 0 & \Sq^1(T) & \Sq^2 (T)\\
    \bZ_2 & 1 & 0 & 0 & \Sq^1(T) & \Sq^2 (T)\\
    \rU(1) &\rU(1)&0 & \bZ_2 & 0 & \bZ_2  \\ \hline
    & 0 & 1 & 2 & 3 & 4 \,\quad,
    \end{array}
\end{equation}
where we have killed the generator $T$ in degree $(2,1)$ and $(2,2)$ as per the discussion after \eqref{superofBZ2}.   We must determine what happens to $\Sq^1(T)$ in $(3,1)$ under $(-1)^{\Sq^2}$.   This amounts to identifying whether $\Sq^2 \Sq ^1 (T)$ is in the kernel of $(-1)^x: \rH^5(\rB\bZ_2\,;\bZ_2) \to \rH^5(\rB\bZ_2\,;\rU(1)) $, where the generators of $\rH^5(\rB\bZ_2\,;\bZ_2)$ are $\{\,\Sq^2 \Sq^1 (T)\,, T \Sq^1(T)\,,\,\\
\Sq^2 \Sq^1 (T)+T \Sq^1(T)\,\}$.
 To discern this we consider the short exact sequence,
\begin{equation}
    \begin{tikzcd}
    0\longrightarrow \bZ_2 \overset{(-1)^x}{\longrightarrow} \rU(1) \overset{x^2}{\longrightarrow}  \rU(1) \longrightarrow  0 \, . \\
\end{tikzcd}
\end{equation}
Let $\Box$ be the Bockstein of this sequence in cohomology such that $\Box: \rH^n(\rB\bZ_2\,;\rU(1)) \to \rH^{n+1}(\rB\bZ_2\,;\bZ_2)$.  Then $\Box (-1)^x = \Sq^1(x)$ and if  $\Sq^1(x)\neq 0 $, that implies $(-1)^x\neq 0$.   Applying $\Sq^1$ to the generators above we have $\Sq^1 \Sq^2 \Sq^1(T) = \Sq^3 \Sq^1 (T) $ by an Adem relation, and $\Sq^3 \Sq^1 (T) = (\Sq^1(T))^2 \neq 0 $ because $\Sq^1(T)$ is in degree three.  Next,
\begin{equation}
\Sq^1 (T\Sq^1(T)) =\Sq^1(T) \Sq^1(T) + \Sq^1 \Sq^1 (T) =(\Sq^1(T))^2. 
\end{equation}
Where we used the fact that $\Sq^1$ acts as a derivation, and  $\Sq^1 \Sq^1 (T) =0 $.  Finally, $\Sq^1(\Sq^2 \Sq^1 (T)+T \Sq^1(T) ) = 0$.  We see that $\Sq^2 \Sq^1 (T)$ is not in the kernel of $(-1)^x$, so that $d_2$ differential is nontrivial.  We find that the $\SH^4(\rB\bZ_2) = \bZ_2$.

We now remark that the $\bZ_4 = \rH^4(\rB\bZ_2\,; \rU(1))$ were the four bosonic braided fusion categories with $\bZ_2$ fusion rules.   We know them explicitly as Vec$_{\bZ_2}$, Semion, anti-Semion, and SVec.  The map to $\SH^4(\rB\bZ_2)$ takes a braided fusion category and tensors it with SVec to produce a braided fusion supercategory.  What we find is that, after this tensor product, SVec becomes equivalent to Vec$_{\bZ_2}$, and the two Semions become equivalent.  This is the statement that $d_2:\rH^3(\rB\bZ_2\,;\bZ_2) \to \rH^5(\rB\bZ_2\,;\rU(1)) $ was nonzero.  The above calculation also says that we do not get any new braided fusion supercategories with $\bZ_2$ fusion rules. 

Another possible way to convince oneself of this is as follows: suppose that there were a class in $\SH^4(\rB\bZ_2)$ with a nontrivial Majorana layer, which is the image of the class on the top row, $\hom(\bZ_2,\bZ_2)$, by the universal coefficient theorem.  This data would imply that the  supercategory had a Majorana object, meaning one with endomorphisms Cliff$_\bC$(1) rather than $\bC$.  The underlying category of our putative braided supercategory is a braided non-super fusion category. Each ordinary object becomes two objects, each Majorana object becomes one object, and the vacuum becomes a boson and a fermion.  That fermion must have fermionic statistics, and also needs to braid trivially with all other objects. Hence, it is an invisible fermion.  If there were a Majorana layer, then the underlying non-super category would have Ising fusion rules.  This is inconsistent with an invisible fermion and so can not happen.  

One could also ponder the existence of a  nontrivial Gu-Wen layer.  If the Gu-Wen layer is nontrivial, then its value is $\Sq^1 t$, where $t$ is the generator of $\rH^\bullet(\rB\bZ_2\,; \bF_2)$ of degree two.  Since $\Sq^1$ is a stable cohomology operator, it commutes with the loop map $ \rH^n(\rB\bZ_2\,; \bF_2) \to \rH^{n-1}(\Omega \rB\bZ_2\,; \bF_2) $.  In our case, $\Omega \rB\bZ_2 =\bZ_2$ with $\Omega$ taking a braided fusion object to its underlying monoidal object.   Let us take $s = \Omega t$ to be the generator of $\rH^\bullet(\Omega \rB\bZ_2\,; \bZ_2 )$, we see $\Sq^1(s) = s^2 \neq 0$.   Thus, if there is a Gu-Wen layer, then it is possible to observe this already on the underlying non-braided category as an effect on the fusion coefficients.  In the presence of a Gu-Wen layer, the underlying non-super category would have $\bZ_4$ fusion rules.  Like the case for the Majorana layer, this is inconsistent with an invisible fermion, and as such can not exist. 

\subsection{Subtheories with SPT}
We return to the last part of section \S\ref{revsymfrac} where the assignment of symmetry fractionalization appears as a choice of zero-form discrete torsion, which we will just call ``discrete torsion", on each subsector of the theory $\mathcal{T}$. The discrete group $\bZ_p$ that we took in \S\ref{defining2groups} does not admit discrete torsion, so for this section we will work with the specific example where $G = \mathbb{Z}_p\times\mathbb{Z}_p$.

We denote the (1+1)$d$ partition function of $\mathcal{T}$ with zero-form and one-form global symmetry placed on a torus as $\cZ_{A_a,A_b,B_a,B_b;\chi^{(2)}}$.  Here, we drop the superscripts on $A$ and $B$ which are background gauge fields for the $G$ symmetry with $(A_a, B_a)\in \mathbb{Z}_p\times\mathbb{Z}_p $.  The subscripts indicate the cycle on the torus which the gauge field wraps. 

The symmetry fractionalization is classified by a class  in $ \rH^2(G;\mathbb{C}[\mathcal{A}_{[1]}]^\times)$ and is part of the information assigned to $\mathcal{T}$.  Due to the fact that $\mathcal{T}$ decomposes into subsectors, the symmetry fractionalization on the full theory was equivalently seen as choices of discrete torsion 
on the subsectors, as given by the homomorphism $f$ at the end of \S\ref{revsymfrac}.  We want to understand how the action of discrete torsion on the subsectors presents itself in the full theory, thereby understanding the action of symmetry fractionalization on $\mathcal{T}$. However, it is expected that acting simply on the subtheories will re-mix in the full theory in a nontrivial way.  In order to understand how discrete torsion acts on the partition function for the full theory, we consider writing the partition function in a basis as
\begin{align}\label{nicebasis}
    \cZ^{\mathcal{T}}_{A_a,X,B_a,Y;\chi^{(2)}} = \int  D A_a D B_b D \xi^{(0)}_{\widehat\kappa} &\exp \left(i\int \chi^{(2)}\cup {\xi^{(0)}_{\widehat\kappa}} \right) \notag\\
    &\times \exp\left(i \int \left( A_b\cup  X + B_b\cup Y \right)\right)  \cZ^{\mathcal{T}/\!/\mathcal{A}_{[1]}}_{A_a,A_b,B_a,B_b;\,{\xi^{(0)}_{\widehat\kappa}} }\,,
\end{align}
where $X$ and $Y$ denote some other background gauge fields for the zero-form symmetry.  The action of discrete torsion by the operator $S_{\widehat{\kappa}}$ on the partition function of the full theory is implemented in this basis to act on the subtheories.  We have
\begin{align}
    S_{{\widehat\kappa}}  \,\cZ^{\mathcal{T}}_{A_a,X,B_a,Y;\chi^{(2)}}&= \int D A_a D B_b D {\xi^{(0)}_{\widehat\kappa}} \exp \left(i\int \chi^{(2)}\cup {\xi^{(0)}_{\widehat\kappa}} \right)   \exp\left(i \int \left( A_b\cup  X + B_b\cup Y \right)\right) \notag \\
    &\hspace{95mm} \times S_{{\widehat\kappa}}  \,\cZ^{\mathcal{T}/\!/\mathcal{A}_{[1]}}_{A_a,A_b,B_a,B_b;{\xi^{(0)}_{\widehat{\kappa}}} }\, \notag\\
    & = \int D A_a D B_b D {\xi^{(0)}_{\widehat\kappa}}\exp \left(i\int \chi^{(2)}\cup {\xi^{(0)}_{\widehat\kappa}} \right)   \exp\left(i \int \left( A_b\cup  X + B_b\cup Y \right)\right) \notag\\
    &\hspace{40 mm}\times \exp\left( i \,\ell_{\xi^{(0)}_{\widehat{\kappa}}} \int \left(A_a\cup B_b-A_b\cup B_a \right) \right) \cZ^{\mathcal{T}/\!/\mathcal{A}_{[1]}}_{A_a,A_b,B_a,B_b; {\xi^{(0)}_{\widehat{\kappa}}}} \notag \\
    & = \int  D A_a D B_b D {\xi^{(0)}_{\widehat{\kappa}}}\exp \left(i\int \chi^{(2)}\cup {\xi^{(0)}_{\widehat{\kappa}}} \right) \notag \\
    &\hspace{10mm}\times \exp \left( i \int \left( A_b\cup\left(X-  \ell_{\xi^{(0)}_{\widehat{\kappa}}}B_a   \right)+B_b\cup \left(Y - \ell_{{\xi^{(0)}_{\widehat{\kappa}}}} A_a\right) \right)\right) \cZ^{\mathcal{T}/\!/\mathcal{A}_{[1]}}_{A_a,A_b,B_a,B_b; {\xi^{(0)}_{\widehat{\kappa}}}}\notag \\
    & = \int D A_a D B_b \exp \left( i \int \left( A_b\cup\left(X- \ell_{\chi^{(2)}_{\widehat{\kappa}}} B_a   \right)+B_b\cup \left(Y -\ell_{\chi^{(2)}_{\widehat{\kappa}}} A_a\right) \right)\right)\notag\\
    &\hspace{90mm}\times\CZ^{\mathcal{T}}_{A_a,A_b,B_a,B_b; \chi^{(2)}} \notag \\
    & = \cZ^{\mathcal{T}}_{A_a,X-\ell_{\chi_{\widehat{\kappa}}} B_a, B_a,Y-\ell_{\chi_{\widehat{\kappa}}} A_a; \chi^{(2)}}\,,
\end{align}
where, $\ell_{{\xi^{(0)}_{\widehat{\kappa}}}}$ is a natural number modulo the order of the one-form symmetry group.

We see that a choice of discrete torsion, when we write $\cZ^{\mathcal{T}}$ in the basis of \eqref{nicebasis}, acts as a permutation matrix if we regard $\mathcal{Z}^\mathcal{T}$ was a vector labeled by its indices.  More precisely, $S_{\widehat{\kappa}}$ acts as a permutation matrix on the $G$ background fields.   This fact is nontrivial when we solely view implementing discrete torsion as a manipulation in 2$d$; however, this becomes clear after coupling to a bulk (2+1)$d$ TFT and interpreting the \textit{topological manipulations} in 2$d$ as permutation of the bulk topological defects \cite{Gaiotto:2020iye}.  By topological manipulations we mean actions which leave any local dynamics the same, but can change the correlators in systems with nontrivial topological sectors.  Our discussion demonstrates that symmetry fractionalization can equivalently be understood from this point of view, with different choices of symmetry fractionalization acting as different permutations of the background zero-form gauge fields.

\subsection{Discrete Torsion and the One-Form Symmetry}

In this subsection we comment on the manipulations within the topological sector of our (1+1)$d$ theory involving the one-form symmetry. In the spirit of the previous section, manipulations such as applying ``discrete one-form torsion", gauging the one-form symmetry, and permuting the local ground states can be understood by coupling to a bulk (2+1)$d$ TFT where the one-form symmetry is a dynamical gerbe \cite{Kapustin:2013uxa}.  In particular, the topological boundary conditions of the defects in the the bulk will give the possible manipulations regarding the topological sectors of the boundary theory.

As a way to interpret the one-form torsion, we can consider taking the subtheories labeled by $\spec(\mathcal{A}_{[1]})(\mathbb C)$ of the discrete $(-1)$-form symmetry and fiber them over a circle so that at discrete points over the circle lives a (1+1)$d$ theory, see figure \ref{fiberedCircle}.  Arranging the families of theories in this way moves us one dimension higher, where the extra dimension involves making the parametrization of the circle into a physical spacetime.   At an energy scale much above the deep IR, operators can act on the spacetime of some subsector and move us between the different sectors.   If however we gauge the one-form symmetry in $\mathcal{T}$ by choosing a projector $\widehat \kappa$ in $\widehat{\mathcal{A}}_{[-1]}$, then flowing to the far IR eliminates the possibility of tunneling out of the subsector labeled by $\widehat{\kappa}$ and then the original theory $\mathcal{T} $ is decomposed into a disjoint union. 
\begin{figure}
    \centering
     \begin{tikzpicture}[thick, scale = .8]
      \def\Depth{4}
        \def\Height{2}
        \def\Width{2}
        \coordinate (O) at (0-2,4,0);
        \coordinate (A) at (0-2,\Width+1+4,0);
        \coordinate (B) at (0-2,\Width+1+4,\Height+1);
        \coordinate (C) at (0-2,0+4,\Height+1);
        \coordinate (D) at (\Depth,0,0);
        \coordinate (E) at (\Depth,\Width,0);
        \coordinate (F) at (\Depth,\Width,\Height);
        \coordinate (G) at (\Depth,0,\Height);
        
         \draw[black,fill=black!10,opacity=0.8] (O) -- (A) -- (B) -- (C) -- cycle;
         \draw[black] (-2,5.5,1.5) node {$\mathcal{T}_{\widehat{1}}$};
     
     \draw[black,dotted] (-2,4,1.33)--(-2,.5,1.33); 
     \tikzstyle{vertex}=[circle,fill=black!70, minimum size=.05pt,inner sep=1.2pt]
    \node[vertex] at (-2,4,1.33) {}; 
    
    \tikzstyle{vertex}=[circle,fill=black!70, minimum size=.05pt,inner sep=1.2pt]
    \node[vertex] at (-2,.5,1.33) {};
     
      \draw[black,fill=black!10,opacity=0.8] (-1,3,0) -- (-1,\Width+1+3,0) -- (-1,\Width+1+3,\Height+1) -- (-1,0+3,\Height+1) -- cycle;
       \draw[black] (-1,4.5,1.5) node {$\mathcal{T}_{\widehat{2}}$};
      
      \draw[black,dotted] (-1,3,1.33)--(-1,0,1.33); 
      \tikzstyle{vertex}=[circle,fill=black!70, minimum size=.05pt,inner sep=1.2pt]
    \node[vertex] at (-1,3,1.33) {}; 
    
    \tikzstyle{vertex}=[circle,fill=black!70, minimum size=.05pt,inner sep=1.2pt]
    \node[vertex] at (-1,0,1.33) {};
      
      \tikzstyle{vertex}=[circle,fill=black!70, minimum size=.05pt,inner sep=1.2pt]
    \node[vertex] at (-1,0,1.33) {};

       \draw[black,fill=black!10,opacity=0.8] (3,4,0) -- (3,\Width+1+4,0) -- (3,\Width+1+4,\Height+1) -- (3,0+4,\Height+1) -- cycle;
        \draw[black,dotted] (3,4,1.33)--(3,.37,1.33); 
        \draw[black] (3,5.5,1.5) node {$\mathcal{T}_{\widehat{M}}$};

         \tikzstyle{vertex}=[circle,fill=black!70, minimum size=.05pt,inner sep=1.2pt]
    \node[vertex] at (3,4,1.33) {}; 
    
    \tikzstyle{vertex}=[circle,fill=black!70, minimum size=.05pt,inner sep=1.2pt]
    \node[vertex] at (3,.37,1.33) {};

         \tikzstyle{vertex}=[circle,fill=black!100, minimum size=.01pt,inner sep=1pt]
    \node[vertex] at (0.5,4.5,1.33) {};
    
    \tikzstyle{vertex}=[circle,fill=black!100, minimum size=.01pt,inner sep=1pt]
    \node[vertex] at (0.8,4.5,1.33) {};
    \tikzstyle{vertex}=[circle,fill=black!100, minimum size=.01pt,inner sep=1pt]
    \node[vertex] at (1.1,4.5,1.33) {};
    \tikzstyle{vertex}=[circle,fill=black!100, minimum size=.01pt,inner sep=1pt]
    \node[vertex] at (1.4,4.5,1.33) {};
       
     \draw[scale=.25, line width = .3mm] \boundellipse{0,0}{10}{2.5};
     \end{tikzpicture}
     \caption{}
     \label{fiberedCircle}
\end{figure}
Suppose now we act by  $\tilde{S}_{{\xi^{(0)}}} $, the one-form discrete torsion operator, on the partition function of $\mathcal{Z}^{\mathcal{T}}$, and then gauge the one-form symmetry with respect to the projector $\widehat{\kappa}$. The operator $\tilde{S}_{{\xi^{(0)}}} $ cups $\xi^{(0)}$ to the connection of the one-form symmetry, and multiplies the partition function by a ``phase".  We give the action of the one-form discrete torsion on the partition function where we suppress the zero-form gauge field indices from \eqref{nicebasis},
\begin{align}
 \int &D \chi^{(2)} \exp\left( -i\,  \int \chi^{(2)}\cup {\delta^{(0)}_{\widehat{\kappa}}} \right) \tilde{S}_{{\xi^{(0)}}} \CZ^{\mathcal{T}}_{\chi^{(2)}} \notag\\
    &=\int D \chi^{(2)} \exp\left( -i\,  \int \chi^{(2)}\cup {\delta^{(0)}_{\widehat{\kappa}}} \right) \exp\left( i\,\int \chi^{(2)} \cup  \xi^{(0)}\, \right) \CZ^{\mathcal{T}}_{\chi^{(2)}}\notag \\
    &= \, \CZ^{\mathcal{T} /\!/ \mathcal A_{[1]}}_{{\xi^{(0)}}-{\delta^{(0)}_{\widehat{\kappa}}}}.
\end{align}

 %
By acting with $\tilde S_{{\xi^{(0)}}}$ on $\CZ^{\mathcal{T}}$ we have shifted the theory to be in the vacuum labeled by ${{\xi^{(0)}}-{\delta^{(0)}_{\widehat{\kappa}}}}$, instead of the vacuum labeled by ${\delta^{(0)}_{\widehat{\kappa}}}$.   We see the one-form discrete torsion acts analogously to the zero-form discrete torsion as a permutation of the $(-1)$-form index, to switch between subsectors.  The subtlety to note is that the permutation of subsectors is not an effect that takes place in the IR, but rather by applying discrete one-form torsion we have modified the projector which gauges the one-form symmetry. 

So far we have performed gauging as a topological manipulation done purely in (1+1)$d$, we can however regard this action as being implemented by a defect in the bulk TFT which filters the family of theories to a particular one. Imposing a Dirichlet boundary condition on the bulk gerbe fields results in the boundary having subsectors, as the one-form symmetry becomes a global symmetry on the boundary. 
We can denote the subsectors as states on the boundary, which can be written as $\ket{\mathcal{T}_{\widehat{i}}}$.

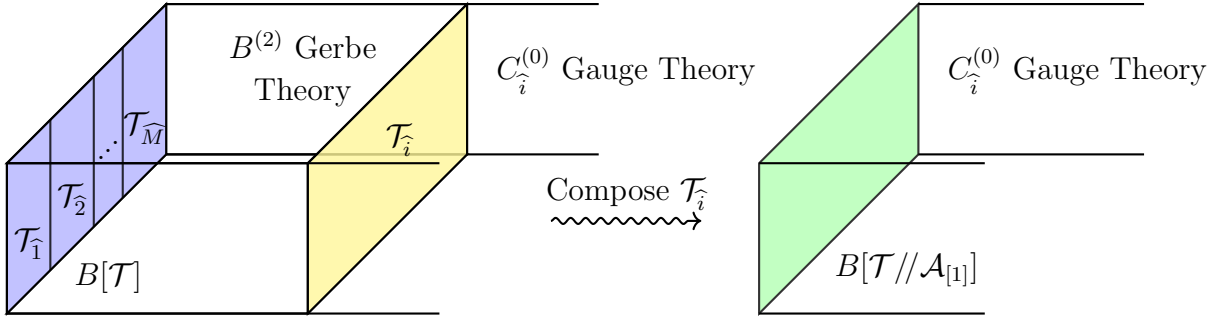
\begin{figure}[ht]
\centering
    \begin{tikzpicture}[thick]
    
        \def\Depth{4}
        \def\DepthTwo{3}
        \def\Height{2}
        \def\Width{2}
        \def\Sep{3}        
        
        \coordinate (O) at (0,0,0);
        \coordinate (A) at (0,\Width,0);
        \coordinate (B) at (0,\Width,\Height+3.5);
        \coordinate (C) at (0,0,\Height+3.5);
        \coordinate (D) at (\Depth,0,0);
        \coordinate (E) at (\Depth,\Width,0);
        \coordinate (F) at (\Depth,\Width,\Height+3.5);
        \coordinate (G) at (\Depth,0,\Height+3.5);
        \draw[black] (O) -- (C) -- (G) -- (D) -- cycle;
        \draw[black] (O) -- (A) -- (E) -- (D) -- cycle;
        \draw[black, fill=blue!30,opacity=0.8] (O) -- (A) -- (B) -- (C) -- cycle;
        \draw[black, fill=yellow!50,opacity=0.8] (D) -- (E) -- (F) -- (G) -- cycle;
        \draw[black] (C) -- (B) -- (F) -- (G) -- cycle;
        \draw[black] (A) -- (B) -- (F) -- (E) -- cycle;
        \draw[below] (\Depth+.25, \Width/2+.7, \Height+1) node{$\mathcal{T}_{\widehat{i}}$};
        \draw[midway] (\Depth/2+.2,\Width-\Width/2+.5
        ,\Height/2) node {\begin{tabular}{c} $B^{(2)}$ Gerbe \\ Theory \end{tabular}};
        \draw[midway] (\Depth/2+\Depth+.5,\Width-\Width/2+.5,\Height/2) node {$C^{(0)}_{\widehat{i}}$ Gauge Theory};
        \coordinate (1) at (0, \Width, 4);
        \coordinate (2) at (0, 0, 4);
        \coordinate (3) at (0, \Width, 2.5);
        \coordinate (4) at (0, 0, 2.5);
        \coordinate (5) at (0, \Width, 1.5);
        \coordinate (6) at (0, 0, 1.5);
         \draw[below] (0, \Width/2, 2.25) node{$.$};
         \draw[below] (0, \Width/2, 2) node{$.$};
         \draw[below] (0, \Width/2, 1.75) node{$.$};
        \draw[black,opacity=.8] (1) -- (2);
        \draw[black,opacity=.8] (3) -- (4);
        \draw[black,opacity=.8] (5) -- (6);
        \draw[below] (0, \Width/2, 4.75) node{$\mathcal{T}_{\widehat{1}}$};
        \draw[below] (0, \Width/2, 3.25) node{$\mathcal{T}_{\widehat{2}}$};
        \draw[below] (0, \Width/2, .75) node{$\mathcal{T}_{\widehat{M}}$};
        \draw[below] (0, -.5, \Height) node{$B[\mathcal{T}]$};
        \draw[below] (\Depth+\DepthTwo+\Sep+1, 0, \Height+1) node{$B[\mathcal{T}/\!/\mathcal{A}_{[1]}]$};
        
        \coordinate (O2) at (\Depth,0,0);
        \coordinate (A2) at (\Depth,\Width,0);
        \coordinate (B2) at (\Depth,\Width,\Height+3.5);
        \coordinate (C2) at (\Depth,0,\Height+3.5);
        \coordinate (D2) at (\Depth+\DepthTwo-1.25,0,0);
        \coordinate (E2) at (\Depth+\DepthTwo-1.25,\Width,0);
        \coordinate (F2) at (\Depth+\DepthTwo-1.25,\Width,\Height+3.5);
        \coordinate (G2) at (\Depth+\DepthTwo-1.25,0,\Height+3.5);
        \draw[black] (O2) -- (D2);
        \draw[black] (A2) -- (E2);
        \draw[black] (B2) -- (F2);
        \draw[black] (C2) -- (G2);
        
        \coordinate (O3) at (\Depth+\DepthTwo+\Sep,0,0);
        \coordinate (A3) at (\Depth+\DepthTwo+\Sep,\Width,0);
        \coordinate (B3) at (\Depth+\DepthTwo+\Sep,\Width,\Height+3.5);
        \coordinate (C3) at (\Depth+\DepthTwo+\Sep,0,\Height+3.5);
        \coordinate (D3) at (\Depth+2*\DepthTwo+\Sep,0,0);
        \coordinate (E3) at (\Depth+2*\DepthTwo+\Sep,\Width,0);
        \coordinate (F3) at (\Depth+2*\DepthTwo+\Sep,\Width,\Height+3.5);
        \coordinate (G3) at (\Depth+2*\DepthTwo+\Sep,0,\Height+3.5);
            \draw[black, fill=green!30,opacity=0.8] (O3) -- (C3) -- (B3) -- (A3) -- cycle;
        \draw[black] (O3) -- (D3);
        \draw[black] (A3) -- (E3);
        \draw[black] (B3) -- (F3);
        \draw[black] (C3) -- (G3);
        \draw[midway] (\Depth/2+\Depth+\DepthTwo+\Sep+.5,\Width-\Width/2+.5,\Height/2) node {$C^{(0)}_{\widehat{i}}$ Gauge Theory};
        \draw[->,decorate,decoration={snake,amplitude=.4mm,segment length=2mm,post length=1mm}] (\Depth+\DepthTwo+\Width/4-2, \Width/2-1.5, \Height/2) -- (\Sep+\Depth+\DepthTwo-\Width/4-2,\Width/2-1.5,\Height/2) node[midway, above] {Compose $\mathcal{T}_{\widehat{i}}$};
        
    \end{tikzpicture}
    \caption{The boundary of the bulk theory is denoted $B[\mathcal{T}]$.  The resulting theories $\mathcal{T}_{\widehat{i}}$ are separated by walls.  When we compose the filter defect in with the boundary, the new boundary becomes that of the bulk theory $\mathcal{T}/\!/\mathcal{A}_{[1]}$.}
        \label{composeBoundaryCond}
\end{figure}

We then place the filter defect such that passing this defect changes the gerbe fields in the bulk to their dual fields, see figure \ref{composeBoundaryCond}.
By composing this defect with the boundary theory, we obtain a composite boundary condition.  Therefore, gauging a one-form symmetry can be treated as implementing this composite boundary condition on the bulk fields of the (2+1)$d$ theory with $C^{(0)}_{\widehat{i}}$ connection.

Another obvious topological manipulation one can also perform is to permute the sectors of the boundary theory, since they are labeled by the one-form symmetry.   From the bulk point of view, this can be seen as a permutation of topological codimension one defects, which end as lines on the boundary.
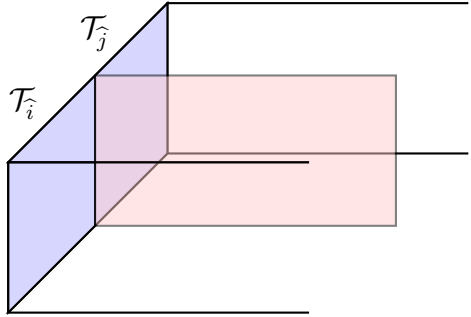
\begin{figure}[ht]
\centering
    \begin{tikzpicture}[thick]
    
        \def\Depth{4}
        \def\DepthTwo{3}
        \def\Height{2}
        \def\Width{2}
        \def\Sep{3}        
        
        \coordinate (O) at (0,0,0);
        \coordinate (A) at (0,\Width,0);
        \coordinate (B) at (0,\Width,\Height+3.5);
        \coordinate (C) at (0,0,\Height+3.5);
        \coordinate (D) at (\Depth,0,0);
        \coordinate (E) at (\Depth,\Width,0);
        \coordinate (F) at (\Depth,\Width,\Height+3.5);
        \coordinate (G) at (\Depth,0,\Height+3.5);
        \draw[black] (O) -- (C) -- (G) ;
        \draw[black] (D)--(O) -- (A) -- (E);
        \draw[black, fill=blue!20,opacity=0.8] (O) -- (A) -- (B) -- (C) -- cycle;
        
        \draw[black] (A) -- (B) -- (F) ;
        \coordinate (3) at (0, \Width, 2.5);
        \coordinate (4) at (0, 0, 2.5);
        \draw[black,opacity=.8] (3) -- (4);
        \coordinate (5) at (\Depth, \Width, 2.5 ) ;
        \coordinate (6) at (\Depth, 0,2.5  );
        \draw[fill = red!20, opacity=.5] (3) --(4) -- (6) -- (5)--cycle;
        \draw[black] (C) -- (B) -- (F);
       \draw[below] (-0.2, \Width+.75, \Height+2.5) node{$\mathcal{T}_{\widehat{i}}$};
        \draw[below] (-0.2, \Width+.75, \Height) node{$\mathcal{T}_{\widehat{j}}$};
        
    \end{tikzpicture}
    \caption{A single defect ending on the boundary separates two subsectors by a line. }
    \label{crossing}
\end{figure}
  The boundary conditions of these defects can be seen as what separates two subsectors of the (1+1)$d$ theory, and thereby crossing a line amounts to traversing between theory $\mathcal{T}_{\widehat{i}}$ and $\mathcal{T}_{\widehat{j}}$, as displayed in figure \ref{crossing}.  The set of ways for these bulk defects to end, and therefore all the ways to traverse between theories, should therefore account for all the ways to permute the boundary theories.  An example of a theory that exhibits the property of being able to move between subsectors as previously described, is a  (1+1)$d$ $\rU(1) $ gauge theory with a charge $q$ massless Dirac fermion \cite{Komargodski:2020mxz}.  One can build a topological local operator $V_k = e^{\frac{2\pi i k}{q}\frac{F_{01}}{e^2}}$ as the symmetry operator of the one-form $\mathbb{Z}_q$ symmetry. We can diagonalize this operator such that it acts on a ground state $\ket{a_1}$ with eigenvalue $e^{\frac{2 \pi i k a_1}{q}}$.  If two states $\ket{a_1}$ and $\ket{a_2}$ are such that $a_1 \not\equiv - a_2 \mod q$ then the following inner product for the overlap between the two states obeys the equality: 
   \begin{equation}
       \bra{a_1} V_k U(t) \ket{a_2} = e^{\frac{-2 \pi i k a_1}{q}} \bra{a_1}U(t) \ket{a_2} = e^{\frac{ 2 \pi i k a_2}{q}}\bra{a_1} U(t) \ket{a_2}\,,
   \end{equation}
   which implies that $\bra{a_1}U(t) \ket{a_2} = 0$.  Here, $U(t)$ is a unitary operator which implements time evolution.  There is no mixing between the subsectors  $\ket{a_1}$ and $\ket{a_2}$, which means the domain walls separating the two sectors have infinite tension.
Another object which we consider is the Wilson line made by a massive probe particle of charge $p \not\equiv 0 \mod q$ which is charged under the one-form symmetry; $W_p = e^{2\pi i p \oint A}$ with $V_k W_p = e^{\frac{2 \pi i k p}{q}} W_p V_k$.  We may therefore allow the Wilson line to surround a subsector specified by a local ground state and calculate 
\begin{align}
    V_k W_p \ket{a_1} &= e^{\frac{2\pi i k p}{q}} W_p V_k \ket{a_1}\notag \\
    & = e^{\frac{2\pi i k (p+a_1)}{q}} W_p \ket{a_1}\,,
\end{align}
which means that the Wilson line separates the different subsectors $\ket{a_1}$ and $\ket{p+a_1}$, because $V_k$ acting on sector $\ket{a_1}$ wrapped with a Wilson line takes us to a different sector $\ket{p+a_1}$.  The Wilson line in this example takes exactly the interpretation as the ending of the bulk defect at the boundary.

\acknowledgments
It is a pleasure to thank Theo Johnson-Freyd  and Justin Kulp for many useful discussions which helped shape this paper, and Jaume Gomis for providing many useful comments on earlier drafts.  We also thank Po-Shen Hsin for his suggestion regarding studying symmetry fractionalization in (1+1)$d$. This research is supported in part by Perimeter Institute for Theoretical Physics. Research at Perimeter Institute is supported in part by the Government of Canada through the Department of Innovation,
Science and Economic Development Canada and by the Province of Ontario through the Ministry of Colleges and Universities.

\appendix
\section{Spectrum of a Ring}
Let us start off with a commutative $K$-algebra denoted $S$, the simplest being the algebra over $\mathbb{C}$. Define $\spec(S) : R \to \hom(S,R)$.  Where $R$ is some set of test objects. If we let $R = \mathbb{C}$ then $\spec(S)(\mathbb{C}) = \hom(S,\mathbb{C})$ taking values in sets, and $\hom$ is as $\mathbb{C}$-linear algebras.  As an example let $S = \mathbb{C}[x]/\left(p(x)=0\right)$ for some arbitrary polynomial $p(x)$ with complex roots.  Then $\spec(S) = \{\lambda \in \mathbb{C} \,|\, (x- \lambda)^{-1}\notin S\}$ i.e. the noninvertible $(x-\lambda)$.  Therefore, given $x$ there exists a homomorphism $\rho : S \to \mathbb{C}$ with $x \mapsto \lambda$. Conversely, given a map $\rho$, then the point $\rho(x)\in \spec(S)$.  In general, the map from $S $ to the $R$ -valued functions on $\spec(S)(R)$ is neither injective nor surjective.  However, if S has finite dimension, is separable, and $R = \mathbb{C}$ then this map is an isomorphism.  This says that $\spec(S)(\mathbb{C})$ in sets determines $S$, which makes $\spec(S)$ into a sheaf as a space. 

\bibliography{symfrac}
\bibliographystyle{JHEP}
\end{document}